\documentclass[fleqn,usenatbib]{mnras}

\usepackage{newtxtext,newtxmath}
\usepackage[T1]{fontenc}
\usepackage{ae,aecompl}

\usepackage{graphicx}	% Including figure files
\usepackage{amsmath}	% Advanced maths commands
\usepackage{subcaption}
\usepackage{pdflscape}
\usepackage{supertabular}

\usepackage{scalerel,tikz}
\usetikzlibrary{svg.path}
\definecolor{orcidlogocol}{HTML}{A6CE39}
\tikzset{orcidlogo/.pic={
 \fill[orcidlogocol] svg{M256,128c0,70.7-57.3,128-128,128C57.3,256,0,198.7,0,128C0,57.3,57.3,0,128,0C198.7,0,256,57.3,256,128z};
 \fill[white] svg{M86.3,186.2H70.9V79.1h15.4v48.4V186.2z}
 svg{M108.9,79.1h41.6c39.6,0,57,28.3,57,53.6c0,27.5-21.5,53.6-56.8,53.6h-41.8V79.1z M124.3,172.4h24.5c34.9,0,42.9-26.5,42.9-39.7c0-21.5-13.7-39.7-43.7-39.7h-23.7V172.4z}
 svg{M88.7,56.8c0,5.5-4.5,10.1-10.1,10.1c-5.6,0-10.1-4.6-10.1-10.1c0-5.6,4.5-10.1,10.1-10.1C84.2,46.7,88.7,51.3,88.7,56.8z};
}}
\newcommand\orcidicon[1]{\href{https://orcid.org/#1}{\mbox{\scalerel*{
\begin{tikzpicture}[yscale=-1,transform shape]
\pic{orcidlogo};
\end{tikzpicture}
}{|}}}}

\usepackage{eso-pic}% http://ctan.org/pkg/eso-pic

\AddToShipoutPictureBG*{%
  \AtPageUpperLeft{%
    \hspace{0.82\paperwidth}%
    \raisebox{-5\baselineskip}{%
      \makebox[0pt][l]{\textnormal{DES-2022-0733}}
}}}%

\AddToShipoutPictureBG*{%
  \AtPageUpperLeft{%
    \hspace{0.73\paperwidth}%
    \raisebox{-6\baselineskip}{%
      \makebox[0pt][l]{\textnormal{FERMILAB-PUB-22-727-PPD}}
}}}%

\title[OzDES H$\beta$ lags]{OzDES Reverberation Mapping Program: H$\beta$ lags from the 6-year survey}

\author[Malik et al.]{
\parbox{\textwidth}{
\Large
U.~Malik$^{1}$\thanks{umang.malik@anu.edu.au},
R.~Sharp$^{1}$\thanks{rob.sharp@anu.edu.au},
A.~Penton$^{2}$,
Z.~Yu$^{3}$,
P.~Martini$^{3,4,5}$,
C.~Lidman$^{1}$,
B.~E.~Tucker$^{1,6,7}$,
T.~M.~Davis$^{2}$,
G.~F.~Lewis$^{8}$,
M.~Aguena$^{9}$,
S.~Allam$^{10}$,
O.~Alves$^{11}$,
F.~Andrade-Oliveira$^{11}$,
J.~Asorey$^{12}$,
D.~Bacon$^{13}$,
E.~Bertin$^{14,15}$,
S.~Bocquet$^{16}$,
D.~Brooks$^{17}$,
D.~L.~Burke$^{18,19}$,
A.~Carnero~Rosell$^{9,20,21}$,
D.~Carollo$^{22}$,
M.~Carrasco~Kind$^{23,24}$,
J.~Carretero$^{25}$,
M.~Costanzi$^{22,26,27}$,
L.~N.~da Costa$^{9}$,
M.~E.~S.~Pereira$^{28}$,
J.~De~Vicente$^{12}$,
S.~Desai$^{29}$,
H.~T.~Diehl$^{10}$,
P.~Doel$^{17}$,
S.~Everett$^{30}$,
I.~Ferrero$^{31}$,
J.~Frieman$^{10,32}$,
J.~Garc\'ia-Bellido$^{33}$,
D.~W.~Gerdes$^{11,34}$,
D.~Gruen$^{16}$,
R.~A.~Gruendl$^{23,24}$,
J.~Gschwend$^{9,35}$,
S.~R.~Hinton$^{2}$,
D.~L.~Hollowood$^{36}$,
K.~Honscheid$^{3,4}$,
D.~J.~James$^{37}$,
K.~Kuehn$^{38,39}$,
J.~L.~Marshall$^{40}$,
J. Mena-Fern{\'a}ndez$^{12}$,
F.~Menanteau$^{23,24}$,
R.~Miquel$^{25,41}$,
R.~L.~C.~Ogando$^{35}$,
A.~Palmese$^{42}$,
F.~Paz-Chinch\'{o}n$^{24,43}$,
A.~Pieres$^{35,44}$,
A.~A.~Plazas~Malag\'on$^{45}$,
M.~Raveri$^{46}$,
M.~Rodriguez-Monroy$^{12}$,
A.~K.~Romer$^{47}$,
E.~Sanchez$^{12}$,
V.~Scarpine$^{10}$,
I.~Sevilla-Noarbe$^{12}$,
M.~Smith$^{48}$,
M.~Soares-Santos$^{11}$,
E.~Suchyta$^{49}$,
M.~E.~C.~Swanson$^{17}$,
G.~Tarle$^{11}$,
G.~Taylor$^{1}$,
D.~L.~Tucker$^{10}$,
N.~Weaverdyck$^{11,50}$,
R.D.~Wilkinson$^{47}$
\\
{\it \footnotesize (Affiliations listed at the end of the paper)}
}
}

\date{Accepted 2023 January 11. Received 2022 December 21; in original form 2022 October 10}

\pubyear{2023}

\begin{document}
\label{firstpage}
\pagerange{\pageref{firstpage}--\pageref{lastpage}}
\maketitle

% Abstract of the paper
\begin{abstract}
% This is a simple template for authors to write new MNRAS papers.
% The abstract should briefly describe the aims, methods, and main results of the paper.
% It should be a single paragraph not more than 250 words (200 words for Letters).
% No references should appear in the abstract.
Reverberation mapping measurements have been used to constrain the relationship between the size of the broad-line region and luminosity of active galactic nuclei (AGN). This $R-L$ relation is used to estimate single-epoch virial black hole masses, and has been proposed for use to standardise AGN to determine cosmological distances. We present reverberation measurements made with H$\beta$ from the six-year Australian Dark Energy Survey (OzDES) Reverberation Mapping Program. We successfully recover reverberation lags for eight AGN at $0.12<z< 0.71$, probing higher redshifts than the bulk of H$\beta$ measurements made to date. Our fit to the $R-L$ relation has a slope of $\alpha=0.41\pm0.03$ and an intrinsic scatter of $\sigma=0.23\pm0.02$ dex. The results from our multi-object spectroscopic survey are consistent with previous measurements made by dedicated source-by-source campaigns, and with the observed dependence on accretion rate. Future surveys, including LSST, TiDES and SDSS-V, which will be revisiting some of our observed fields, will be able to build on the results of our first-generation multi-object reverberation mapping survey.
\end{abstract}

% Select between one and six entries from the list of approved keywords.
% Don't make up new ones.

%%%%%%%%%%%%%%%%%%%%%%%%%%%%%%%%%%%%%%%%%%%%%%%%%%

%%%%%%%%%%%%%%%%% BODY OF PAPER %%%%%%%%%%%%%%%%%%

\section{Introduction}

Reverberation Mapping (RM) has been established as the leading technique for direct determination of black hole masses ($M_{\rm BH}$) in active galactic nuclei (AGN) outside of the local Universe. Reverberation measurements anchor the scaling relations used to estimate single-epoch virial BH masses \citep{Vestergaard2002,Vestergaard2006}. Over the past decade, RM programs conducted by the Australian Dark Energy Survey (OzDES) and Sloan Digital Sky Survey (SDSS) have leveraged high-multiplexed spectroscopy to perform RM on an `industrial' scale, aiming to increase the number of measurements available by an order of magnitude \citep{King2015,Shen2015}. These programs have resulted in over one hundred new lag measurements, while also identifying and addressing the complexities of performing RM on such large scales. 

As reverberation mapping resolves the innermost regions of AGN in the time-domain, rather than spatially, it allows the study of these regions out to high redshifts. RM utilises the difference in the light travel time between the variation of the continuum emission from the accretion disk around the central supermassive black hole (SMBH) and the reprocessed emission from the photoionised broad-line region (BLR) \citep{Blandford1982,Peterson1993}. The time-delay, $\tau$, can be measured using multi-epoch photometry and spectroscopy to probe the continuum and BLR response respectively in order to infer the radius of the BLR ($R_{\rm BLR} = c\tau$). Together with the BLR velocity dispersion, $\Delta V^2$, inferred from the width of the emission line, this can be used to estimate the mass of the SMBH using the virial product:
\begin{equation}\label{virial}
    M_{\textrm{BH}} = f\frac{R_{\textrm{BLR}} \Delta V^2}{G}
\end{equation}
 The geometry, orientation, and kinematics of the BLR are encapsulated by the dimensionless scale factor $f$, which is calibrated using sources with independent measurements from RM and the $M_{\rm BH}-\sigma_*$ relation \citep{Ferrarese2000,Gebhardt2000,Onken2004,Woo2015}.
 
Due to the intensity of observational resources required, early programs targeted a small number of local AGN with campaign durations of less than one year. To achieve the fidelity required for RM, they observed bright, highly varying AGN. As a result, the targeted samples are typically biased towards AGN in the local Universe ($z<0.3$). Lag measurements were made for 63 AGN with the H$\beta$ line \citep[e.g.,][]{Peterson1998,Kaspi2000,Peterson2004,Bentz2009}. These measurements were used to constrain the relationship between the AGN luminosity and the radius of the BLR ($R-L$ relation), which exhibited relatively low intrinsic scatter, with a slope consistent with that expected by photoionisation physics \citep{Bentz2009,Bentz2013}. This relation calibrates secondary mass-scaling relations used to estimate single-epoch virial BH masses for samples of thousands of AGN \citep[e.g. SDSS DR7 quasar catalogue,][]{Shen2011}. The $R-L$ relation has also been proposed as a way to standardise AGN for use as luminosity distance indicators for cosmology \citep{Watson2011,MartinezAldama2019,Khadka2022}.

These BH mass estimates have a significant $\sim$0.5 dex uncertainty, due to our limited understanding of BLR geometry and kinematics and the small sample of reverberation measurements, among other factors \citep{Shen2013}. The former issue is addressed by  conducting observationally intensive velocity-resolved RM \citep[e.g.][]{Bentz2010,Grier2013,Pancoast2014b,Du2018,U2022}, and using dynamical modelling methods \citep[e.g. CARAMEL,][]{Pancoast2014a}; as well as spectroastrometry of the BLR in local AGN \citep[e.g. GRAVITY,][]{GRAVITY2018}. Other programs are targeting a more diverse range of sources. The super-Eddington accreting massive black holes (SEAMBH) program has observed over 40 luminous AGN \citep{Du2014,Wang2014,Du2015,Du2016,Du2018,Hu2021}. The updated $R-L$ relation using these H$\beta$ measurements has increased intrinsic scatter, and an observed dependence on accretion rate. However, these programs are still only targeting AGN at $z<0.4$.  

The SDSS-RM Project and our OzDES-RM Program have pioneered RM on an `industrial-scale', observing hundreds of AGN probing a wide range of AGN luminosities and redshifts. These programs have added over 100 new lag measurements, and have enabled the Mg\,\textsc{ii} and C\,\textsc{iv} $R-L$ relations to be constrained with statistically significant samples \citep[][Penton et al. in prep]{Grier2017,Hoormann2019,Grier2019,Homayouni2020,Yu2021,Yu2022}. The first generation of multi-object RM surveys have however highlighted problems with monitoring hundreds of targets. This includes challenges both with data quality \citep[signal-to-noise, limited temporal coverage;][]{Malik2022} as well as with reverberation lag recovery techniques and biases such as aliasing \citep{Li2019,Penton2022}. 

OzDES focused mainly on high redshift AGN; however, about 10\% of the sample contains H$\beta$ emission within the spectroscopic window. This will allow comparison of our OzDES measurements with the large sample of existing H$\beta$ measurements, in order to examine the consistency of multi-object RM with earlier dedicated source-by-source observations. We present the H$\beta$ lag results from our 6-year survey. Section \S\ref{sec:data} details the observations obtained by OzDES and the data calibration procedures. In Section \S\ref{sec:lag_rec} we describe the techniques we used for lag recovery and the selection criteria we apply to define our final sample. In Section \S\ref{sec:results} we present our successful lag measurements and black hole masses, as well as an updated $R-L$ relationship, and discuss these results in \S\ref{sec:discussion}. We summarize our results and then present an outlook for the future work in Section \S\ref{sec:summary}. Throughout this work we adopt a flat $\Lambda$CDM cosmology, with $\Omega_{\Lambda}=0.7$, $\Omega_{M}=0.3$, and $H_0=70$ km\,s$^{-1}$\,Mpc$^{-1}$.

\section{Data}
\label{sec:data}
The Australian Dark Energy Survey (OzDES) provided follow-up spectroscopic observations of the 10 supernova fields observed by the Dark Energy Survey (DES). The DES supernova fields are located in the ELAIS, XMM-Large Scale Structure, Chandra deep-field South, and SDSS Stripe 82 regions \citep{Kessler2015,Morganson2018}. These fields were observed in the $griz$ filters with the Dark Energy Camera (DECam) on the 4-metre Blanco telescope at Cerro Tololo Inter-American Observatory (CTIO) \citep{Flaugher2015}. From 2013 to early 2018, the fields were observed with $\sim$6 day cadence over a 5-6 month season (August to January), with additional science verification data taken in late 2012 to early 2013, and additional data taken on a monthly cadence in late 2018. OzDES conducted follow-up multi-object spectroscopic observations with the 2dF multi-object fibre positioning system and the AAOmega spectrograph \citep[3700-8800\,\AA, ][]{Sharp2006} on the 3.9-metre Anglo-Australian Telescope (AAT) \citep{Yuan2015,Childress2017,Lidman2020}. The OzDES observations were made over the same 5-6 month season with approximately monthly cadence from 2013 to 2019.

\subsection{Target sample}

\begin{figure}
	\centering
	\includegraphics[width=\columnwidth]{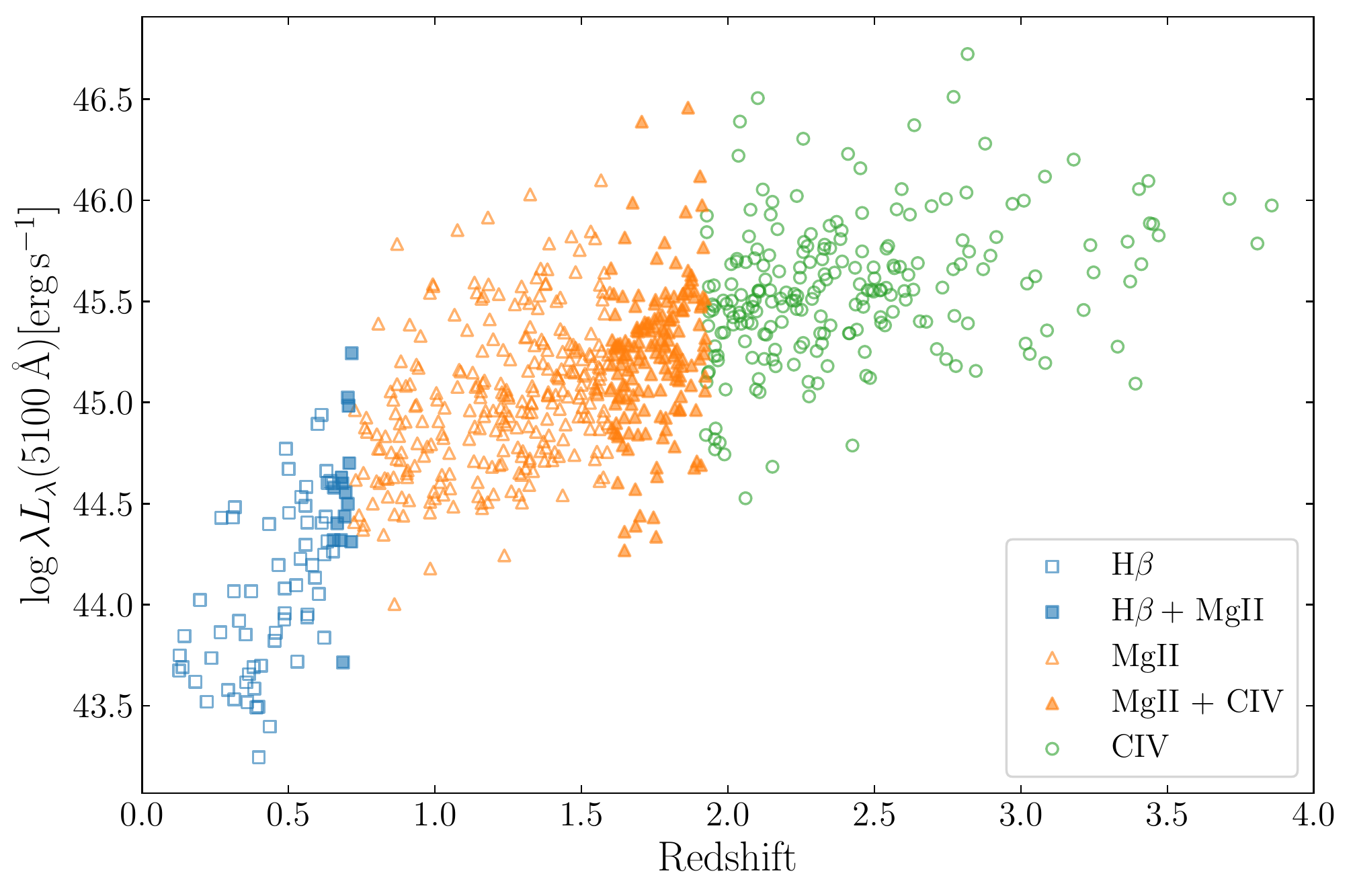}
	\caption{Distribution of redshifts and monochromatic luminosities at 5100\,\AA\ for the 735 AGN in the OzDES RM sample. The H$\beta$ sample extends to $z=0.75$, with 15 sources overlapping with our Mg\,\textsc{ii} sample. } 
    \label{fig:ozdes_AGN}
\end{figure}

\begin{figure}
    \centering
    \includegraphics[width=\columnwidth]{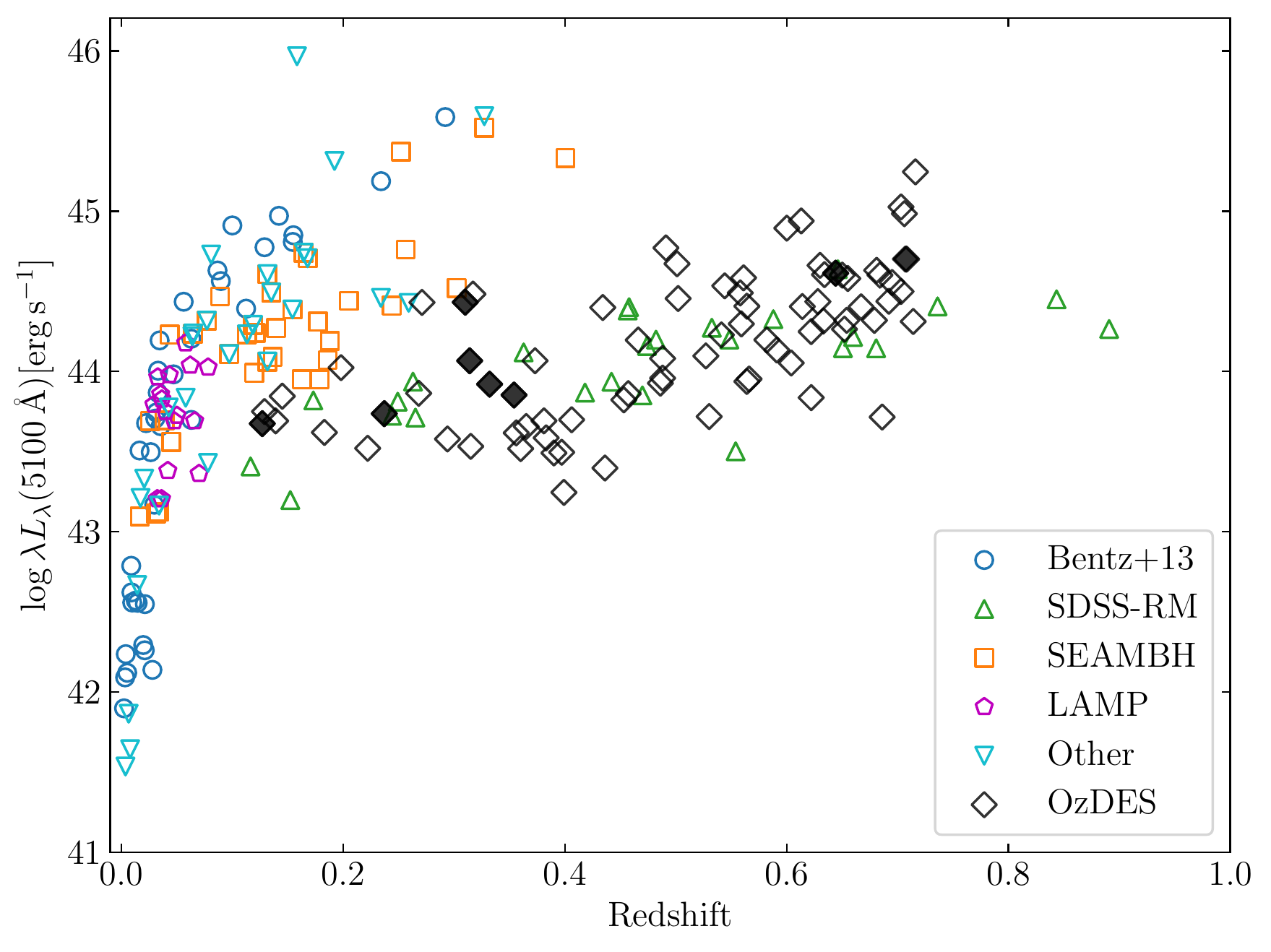}
    \caption{The redshift and luminosity distribution for the OzDES H$\beta$ sample (open diamonds), final sample (filled diamonds, see \S\ref{sec:results}), and existing measurements from \citet[][and references therein]{Bentz2013}; SDSS-RM \citep[][quality 4 and 5]{Grier2017}; SEAMBH  \citep{Du2014,Wang2014,Du2015,Du2016,Du2018,Hu2021}; Lick AGN Monitoring Project \citep[LAMP,][]{U2022}; and other measurements from \citet{Bentz2009,Barth2013,Bentz2014,Pei2014,Lu2016,Bentz2016a,Bentz2016b,Fausnaugh2017,Zhang2019,Rakshit2019,Li2021}, of which measurements published before 2019 are compiled by \citet{MartinezAldama2019}. }
    \label{fig:z_v_lum}
\end{figure}

After completing the final data reduction for our survey, the OzDES RM Program sample comprises 735 AGN (reduced from an initial sample of 771 due to a change in the location of the DECam inter-chip gaps between survey definition and campaign observations), with redshifts ranging up to $z\sim4$, and with apparent magnitudes 17.2 $<$ $r_{\rm{AB}}$ $<$ 22.3 mag. The redshift and luminosity distribution of these AGN is shown in \autoref{fig:ozdes_AGN}.

Of these 735 AGN, the H$\beta$ line and nearby continuum falls within the AAOmega spectral range for 78 sources. The expected observer-frame lags for the H$\beta$ sample (governed by luminosity and time-dilation) range from $\sim$20 days to just over 200 days. In previous analyses, our emission-line flux measurements were made using spectra that were co-added by observing run \citep{Hoormann2019} (typically 4-7 days during dark time each month). However, some fields were observed over multiple nights within a single observing run. To maximise the cadence of our sampling we treated the spectra obtained on different nights as separate epochs for our emission-line light curves. This is particularly valuable for rapidly reverberating sources. The redshift and luminosity range of our H$\beta$ sample, and H$\beta$ measurements from the literature are shown in \autoref{fig:z_v_lum}. 

We do not measure the continuum luminosity directly from the spectra due to fibre aperture effects from variable atmospheric seeing and fibre placement uncertainties. From the average $r$-band magnitude and redshift of the AGN, we estimated the monochromatic continuum flux at 5100\,\AA\ using the DECam $r$-band filter transmission curve and the SDSS quasar template \citep{VandenBerk2001}. The template is scaled to the magnitude of the source, assuming $L_{\rm bol}=9\, \lambda L_{\lambda}$\,(5100\,\AA) \citep{Kaspi2000}. 

\subsection{Flux calibration and measurements}

The DES photometry is calibrated consistently through to Y6 using the DES data reduction pipeline \citep{Morganson2018,Burke2018}. We perform a spectrophotometric flux calibration and line flux measurement following \citet{Hoormann2019}. The local continuum windows for continuum subtraction are 4760 to 4790\,\AA\ and 5100 to 5130\,\AA. The calibration uncertainties of the line flux were measured using the F-star warping function method as detailed in \citet{Yu2021}.

\section{Lag recovery and reliability}\label{sec:lag_rec}

\subsection{Time series analysis}

To measure the reverberation lags for our sample we use the interpolated cross-correlation function \citep[ICCF;][]{Gaskell1987} and \textsc{JAVELIN} \citep{Zu2011,Zu2013} methodologies. The \textsc{JAVELIN} method models the AGN continuum variability using a damped random walk \citep[DRW;][]{Kelly2009,Kozlowski2010,MacLeod2010}. It assumes the emission-line light curve is a scaled, smoothed and shifted version of the continuum light curve, that can be described by the convolution of the continuum with a top-hat transfer function. Using Markov Chain Monte Carlo (MCMC), it constrains the characteristic variability amplitude and damping timescale of the continuum light curve. Applying this DRW fit as a prior, it simultaneously fits both the continuum and emission-line light curves, to derive the posterior distributions of the transfer function parameters: lag, top-hat width and scale factor, as well as an updated DRW amplitude and timescale. We allow these parameters to vary freely, while setting a lag prior of [-3$\tau_{\rm exp}, 3\tau_{\rm exp}$], where $\tau_{\rm exp}$ is the expected H$\beta$ lag for the source using the \citet{Bentz2013} $R-L$ relation. 

We use the \textsc{PyCCF} code to perform the ICCF method \citep{2018ascl.soft05032S}. This linearly interpolates the continuum and emission-line light curves over a user defined grid spacing, and cross-correlates the interpolated light curves as a function of time-lag. The centroid of the cross-correlation function (CCF) is computed as the median of the CCF values $>0.8r_{\rm max}$ counted out from the peak of the CCF, $r_{\rm max}$. A cross-correlation centroid distribution (CCCD) is obtained from 10,000 Monte Carlo realisations of the flux randomisation and random subset sampling (FR/RSS) process \citep{Peterson1998}, which accounts for the flux measurement uncertainties and potentially spurious correlations between light curve points. Following \citet{Hoormann2019}, we set the interpolation grid spacing to 3 days, and the $r_{\textrm{max}}$ threshold to 0.5. We use the same lag prior as for \textsc{JAVELIN}, from [-3$\tau_{\rm exp}, 3\tau_{\rm exp}$]. 

The recovered lag, $\tau$, with lower and upper uncertainties, $\sigma_{\tau}$, are taken to be median and 16th and 84th percentiles of the lag probability distribution functions (PDF) from \textsc{JAVELIN} and \textsc{PyCCF}. 

\subsection{Null hypothesis test}

Our survey window function for the (expected) short H$\beta$ lags is less than ideal and the signal-to-noise of the spectroscopy is only modest. We wish to test the null hypothesis that our lag recovery is not simply a product of the interaction of the window function with underlying red-noise correlation in the photometric light curves (an underlying assumption of any RM technique). Therefore, we do not report extensive light curve simulations of the H$\beta$ sample \citep[as per][]{U2022}, as this would simply reuse the same window function properties with the added uncertainty of the appropriateness of the variability model for our sources. Instead we randomised the spectroscopic light curves (with flux values shuffled while retaining the dates of observation), and cross correlated this with the original photometric light curve. We find the $r_{\rm max}$ value of the cross correlation coefficient $r$ at the peak of the CCF, and compare it to that found from the CCF of the original light curves. After 1000 iterations, the $p$-value was calculated as the fraction of $r_{\rm max}$ values which exceeded the original $r_{\rm max}$. 

By randomising the emission-line light curves we generate uncorrelated light curves that do not posses a reverberation signature. We chose to randomise the observed spectroscopic light curve, rather than using simulated light curve realisations, as done by \citet{U2022}. Simulated light curve realisations have the same variability timescale and underlying lag as the source, and therefore may result in significant spurious correlations at random lag values.  

\begin{figure}
    \centering
    \includegraphics[width=\columnwidth]{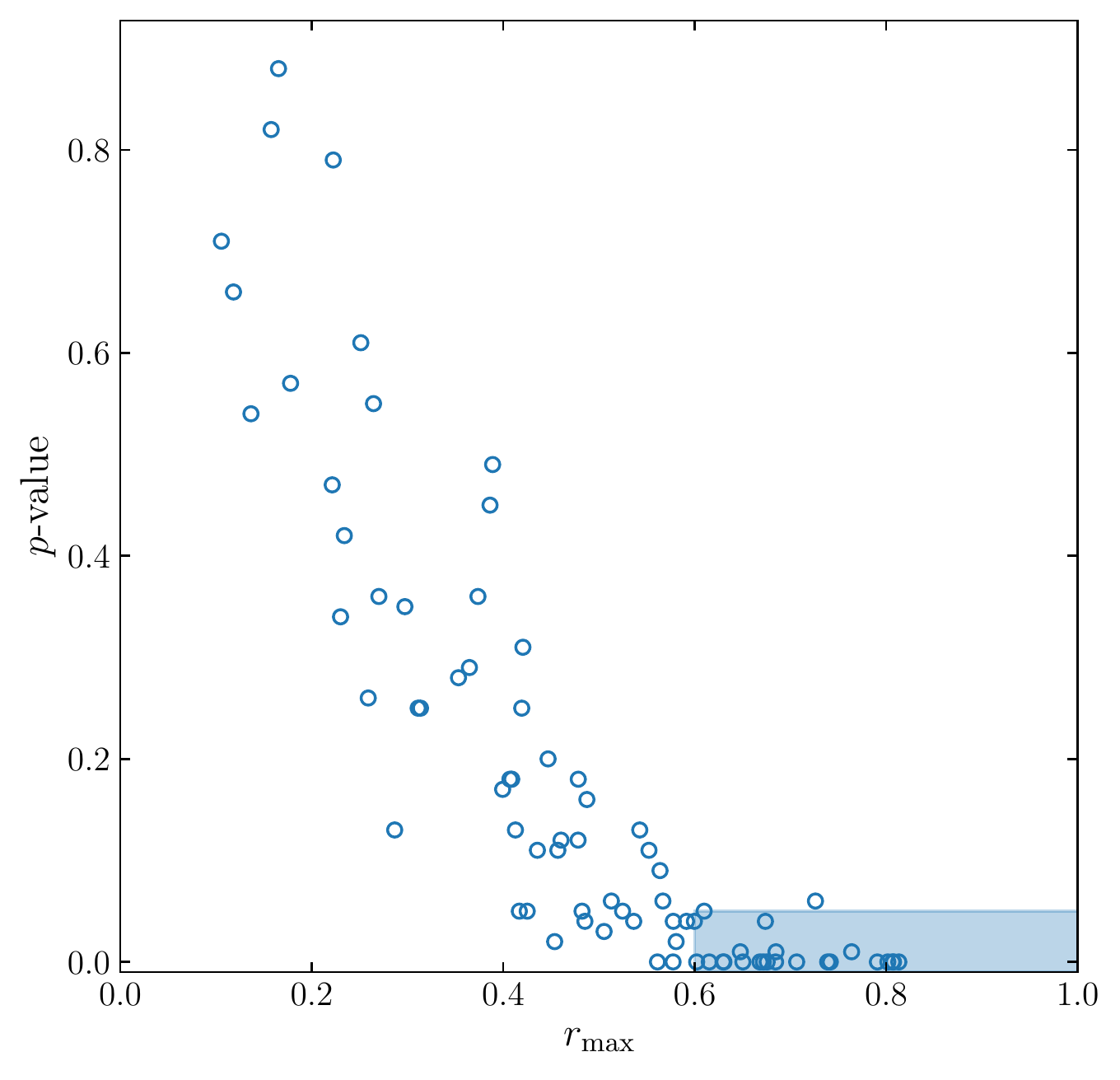}
    \caption{$p$-value vs. $r_{\rm max}$ for our 78 AGN. The $r_{\rm max}$ is of the original pair of light curves. The shaded region indicates our selection criteria requiring a $p$-value $< 0.05$ and $r_{\rm max} > 0.6$.}
    \label{fig:ptest}
\end{figure}

Figure \ref{fig:ptest} shows the results of this test applied to each of the H$\beta$ sources. We see that below a $r_{\rm max}$ of 0.6, there are always a significant number of uncorrelated signals which exceed this $r_{\rm max}$, resulting in high $p$-values. Therefore we can not trust a result which has a $r_{\rm max}$ below 0.6.

\subsection{Selection criteria} \label{sec:criteria}

Based on the results of simulations from \citet{Li2019}, \citet{Yu2020} and \citet{Penton2022}, we adopt the lag and uncertainties from \textsc{JAVELIN} as the final lag. We define a successful lag recovery as meeting the following criteria:

\begin{itemize}
    \item The upper and lower lag uncertainties $\sigma_{\tau,{\rm JAV}}$ are less than $|\tau_{\rm JAV}|$, or 30 days, whichever is greater
    \item $\tau_{\rm PyCCF}$ lies within the 2$\sigma_{\tau,{\rm JAV}}$ uncertainties
    \item $p$-value $< 0.05$ and $r_{\rm max} > 0.6$
\end{itemize}

As the expected lags of our sources range from $\sim$20 to 200d, we use a relative rather than absolute cut on the lag uncertainties. We require the lags recovered from \textsc{JAVELIN} and \textsc{PyCCF} to agree in an attempt to exclude cases where PDF's are flat or have significant aliasing peaks. Our $p$-test and $r_{\rm max}$ criteria ensure there is significant correlation between the light curves that is unlikely to be spurious.

\section{Results} \label{sec:results}

\renewcommand{\arraystretch}{1.4}
\begin{table*}
\captionsetup{justification=centering}
\caption{Results for our final sample of eight AGN. Columns left to right: DES name (J2000), redshift, observer-frame \textsc{JAVELIN} lag, observer-frame \textsc{PyCCF} lag, $p$-value, monochromatic luminosity at 5100\AA, line dispersion measured from mean spectrum, virial black hole mass, and dimensionless accretion rate.  }
\begin{tabular}{ccccccccc} \hline
Source     & $z$     & $\tau_{\rm JAV}$          & $\tau_{\rm PyCCF}$             & $p$-value & $\log({\lambda}L_{5100})$ & $\sigma_{\rm mean}$ & $M_{\rm BH} $ & $\dot{M}$ \\
 &  & (days) & (days) &  & (erg\,s$^{-1}$) & (km\,s$^{-1}$) & ($\times 10^7 M_{\sun}$) & \\ \hline
DES J002802.42-424913.52 & 0.127 & $18^{+3}_{-3}$ & $16^{+10}_{-18}$ & 0.000 & 43.67 & $1393\pm3$ & $2.64^{+0.85}_{-0.87}$ & 1.43 \\
DES J024347.34-005354.84 & 0.237 & $34^{+1}_{-2}$ & $32^{+19}_{-13}$ & 0.000 & 43.74 & $1658\pm6$ & $6.62^{+1.87}_{-1.89}$ & 0.28 \\
DES J022330.16-054758.06 & 0.354 & $22^{+27}_{-20}$ & $21^{+34}_{-47}$ & 0.002 & 43.85 & $1846\pm11$ & $4.86^{+6.12}_{-4.62}$ & 0.79 \\
DES J002904.43-425243.04 & 0.644 & $98^{+27}_{-57}$ & $87^{+31}_{-177}$ & 0.002 & 44.61 & $1851\pm4$ & $17.8^{+7.0}_{-11.4}$ & 0.80 \\
DES J022617.85-043108.99 & 0.707 & $67^{+49}_{-18}$ & $33^{+73}_{-80}$ & 0.006 & 44.70 & $1691\pm8$ & $9.78^{+7.60}_{-3.76}$ & 3.63 \\ \hline
DES J034028.46-292902.41 & 0.310 & $51^{+9}_{-6}$ & $54^{+24}_{-22}$ & 0.002 & 44.43 & $1732\pm5$ & $10.3^{+3.3}_{-3.1}$ & 1.30 \\
DES J022249.67-051453.01 & 0.314 & $25^{+7}_{-5}$ & $29^{+45}_{-55}$ & 0.005 & 44.07 & $1883\pm5$ & $5.99^{+2.28}_{-2.00}$ & 1.08 \\
DES J003954.13-440509.97 & 0.332 & $48^{+12}_{-7}$ & $56^{+17}_{-37}$ & 0.001 & 43.92 & $1856\pm11$ & $10.9^{+4.1}_{-3.4}$ & 0.20 \\ \hline
\end{tabular}
\label{tab:results}
\end{table*}

\begin{figure*}
    \captionsetup[subfigure]{slc=off}
    \centering
    \begin{subfigure}[b]{\textwidth}
       {\includegraphics[width=1\linewidth]{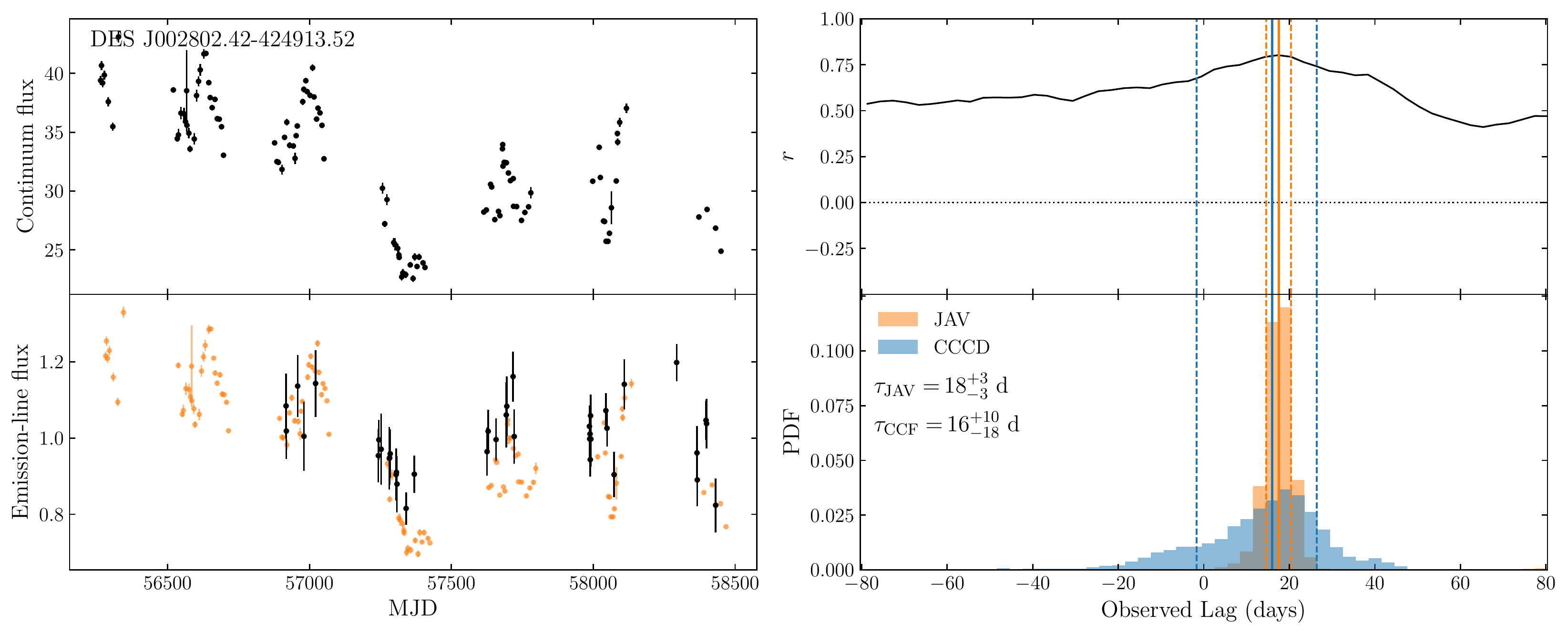}}
    \end{subfigure}
    
    \begin{subfigure}[b]{\textwidth}
       \includegraphics[width=1\linewidth]{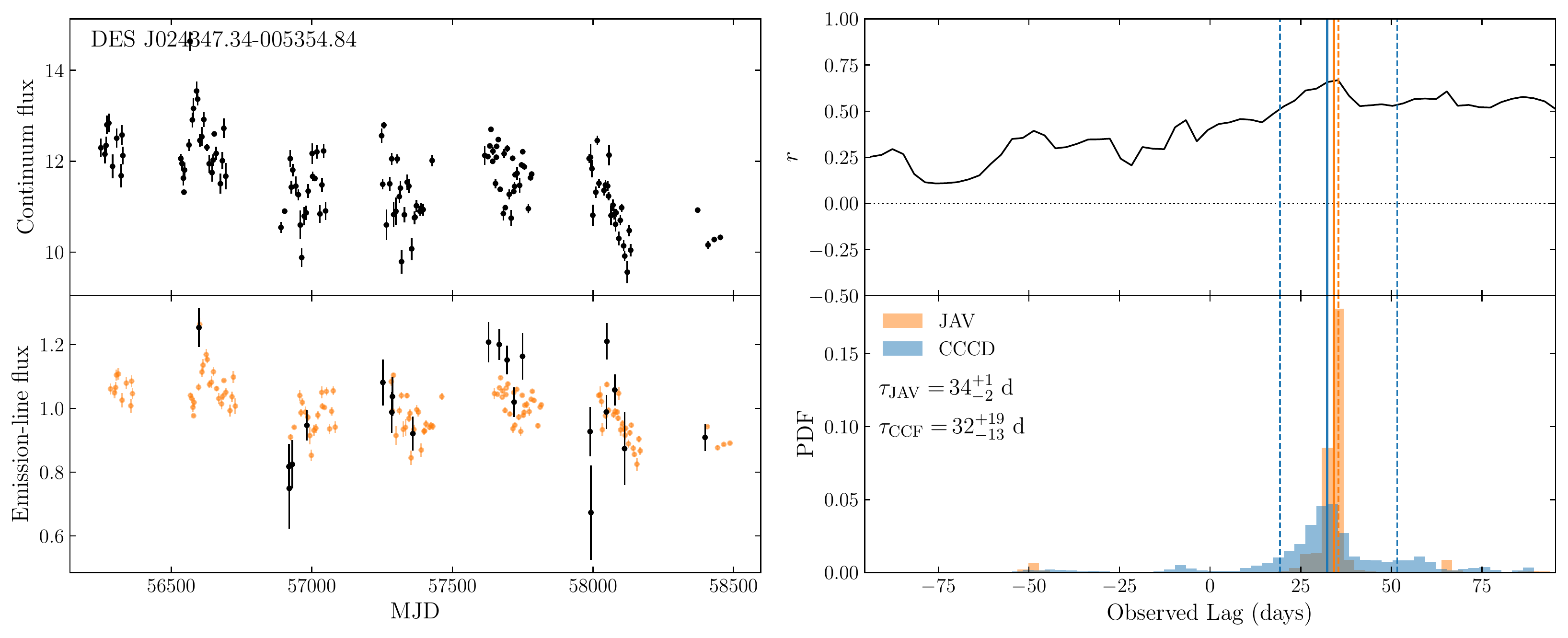}
    \end{subfigure}
    
    \begin{subfigure}[b]{\textwidth}
       \includegraphics[width=1\linewidth]{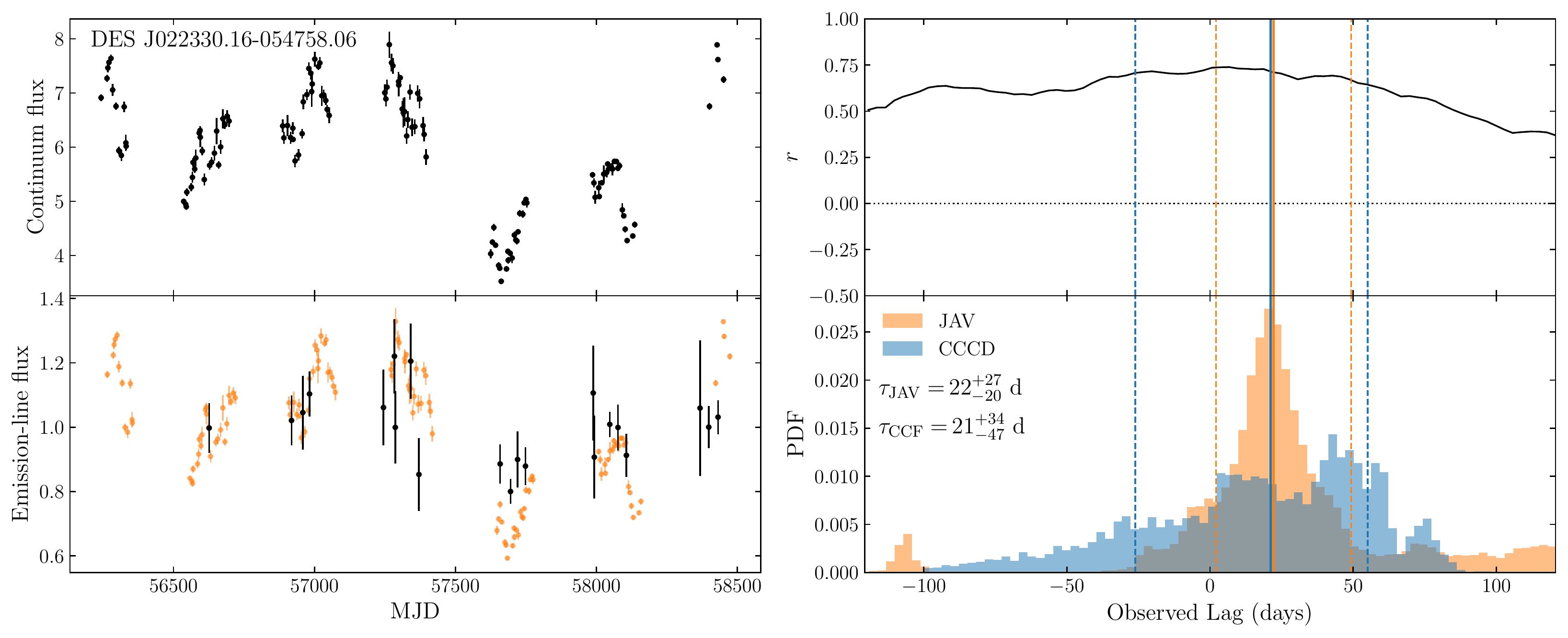}
    \end{subfigure}

    \caption{For each source, the upper-left panel shows the DECam $g$-band continuum light curve, and lower-left panel the H$\beta$ emission-line light curve, with the continuum light curve phase-shifted by the \textsc{JAVELIN} lag. The fluxes have been re-scaled. The upper-right panel shows the cross-correlation function computed using \textsc{PyCCF}. The lower-right panel shows the cross-correlation centroid distribution (CCCD) from \textsc{PyCCF} in blue, and the lag posterior from \textsc{JAVELIN} in orange, with the vertical solid and dashed lines indicating the recovered lag and upper and lower uncertainties from each method.}
    \label{fig:lags1}
\end{figure*}

\begin{figure*}
    \captionsetup[subfigure]{slc=off}
    \centering
    \begin{subfigure}[b]{\textwidth}
       \includegraphics[width=1\linewidth]{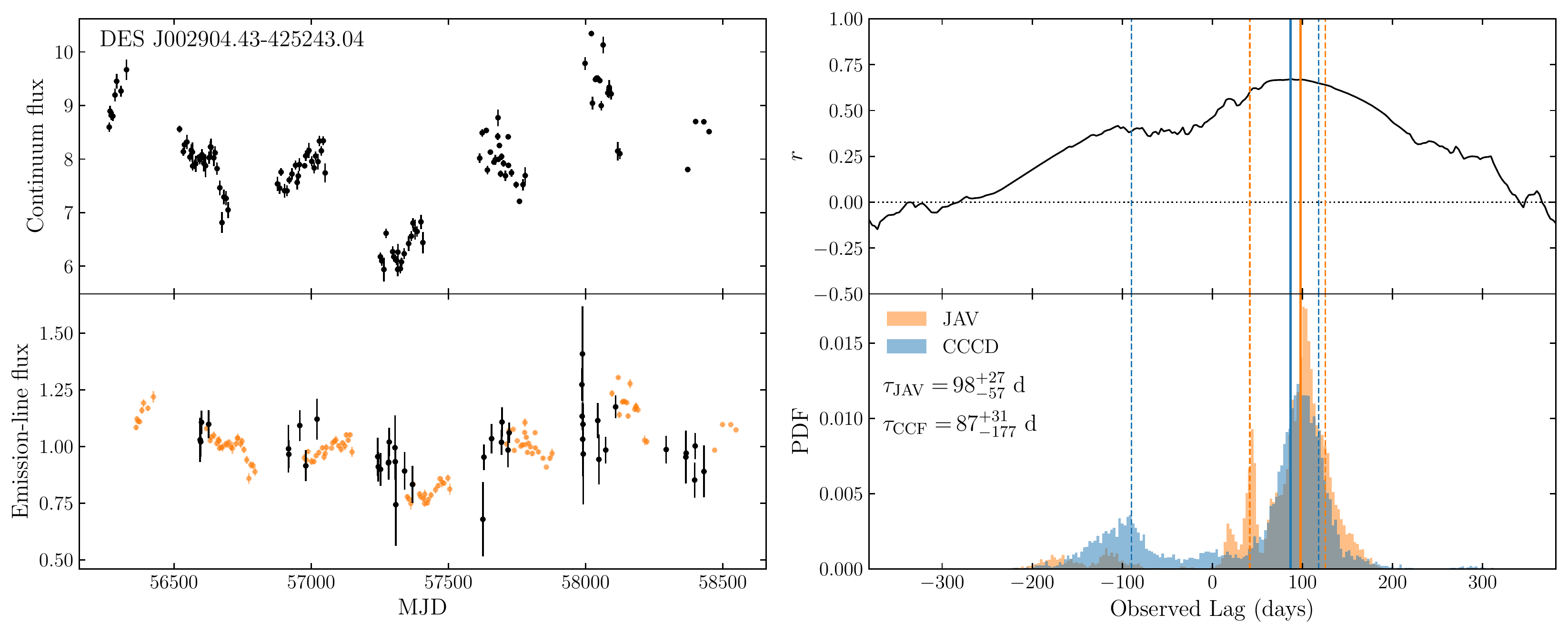}
    \end{subfigure}
    
    \begin{subfigure}[b]{\textwidth}
       \includegraphics[width=1\linewidth]{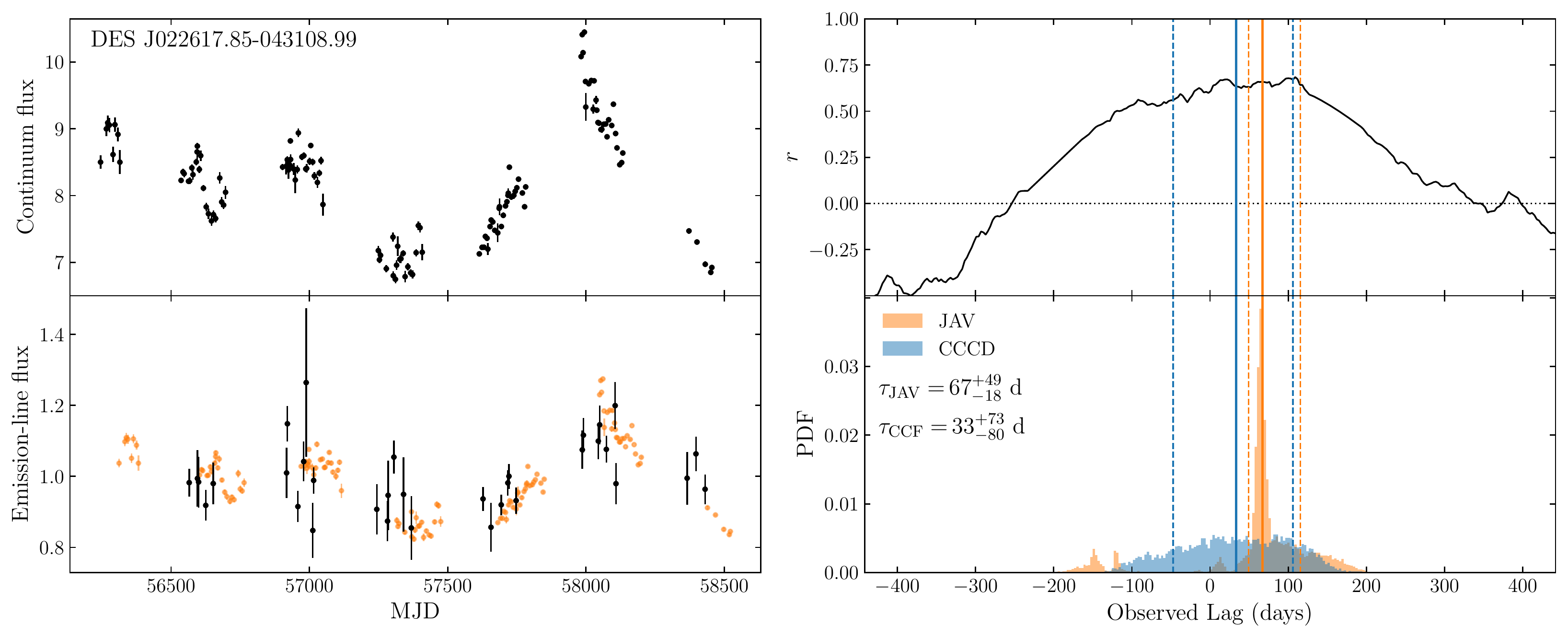}
    \end{subfigure}
    
        \caption{Same as for \autoref{fig:lags1}.}
    \label{fig:lags2}
\end{figure*}

We successfully measure lags for five H$\beta$ sources with the criteria listed in \S\ref{sec:criteria}. The recovered lags are given in \autoref{tab:results}. We present the light curves, and the lag distributions for each of the five sources in \autoref{fig:lags1} and \autoref{fig:lags2}. The re-scaled and phase-shifted light curves for each of the sources display visually discernible reverberation. For each source, there is overlap between the phase-shifted light curves. In most cases, the lag PDF's from \textsc{PyCCF} and \textsc{JAVELIN} agree, although there is significant scatter in the \textsc{PyCCF} CCCD's for some sources. However, even for these cases, the \textsc{JAVELIN} lag is well defined, which is consistent with the finding that \textsc{JAVELIN} is better able to recover short lags than \textsc{PyCCF} \citep{Li2019}.

\subsection{Additional sources} \label{sec:jav_aliasing}
\begin{figure*}
    \captionsetup[subfigure]{slc=off}
    \centering
    \begin{subfigure}[b]{\textwidth}
       \includegraphics[width=1\linewidth]{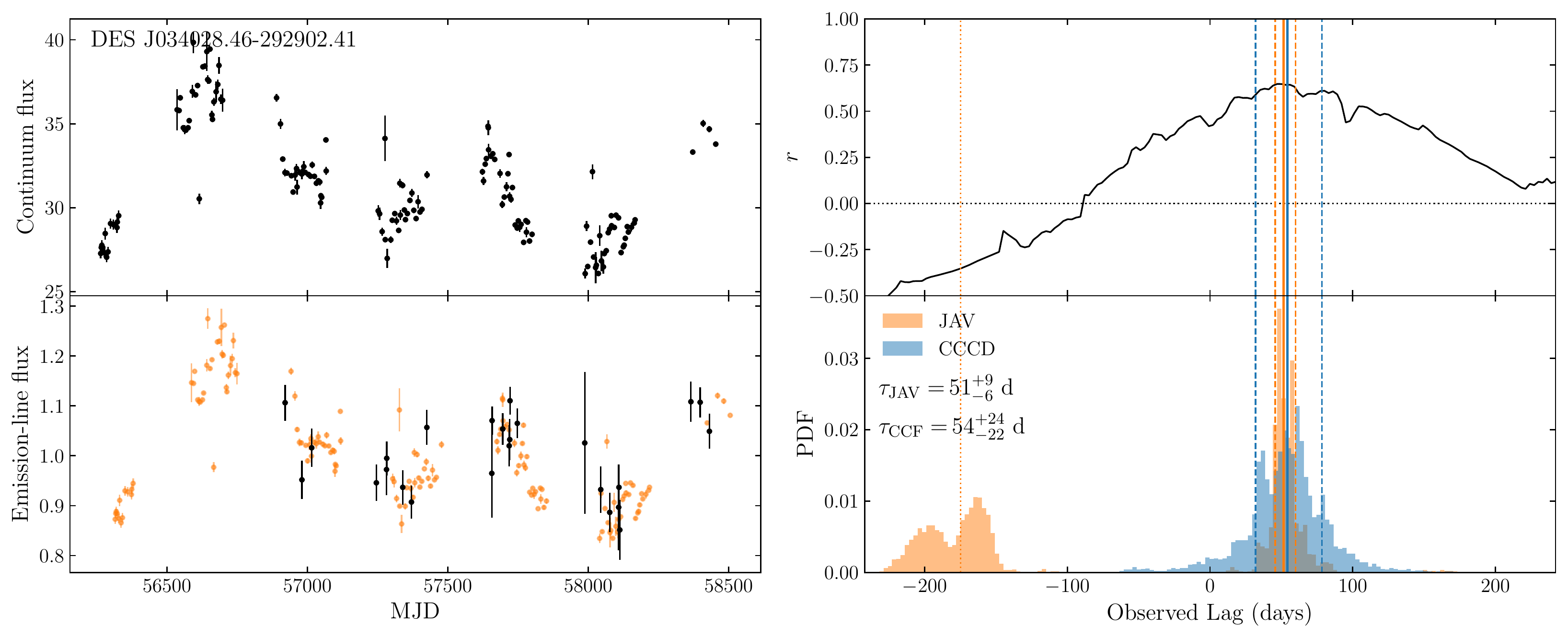}
    \end{subfigure}
    
    \begin{subfigure}[b]{\textwidth}
       \includegraphics[width=1\linewidth]{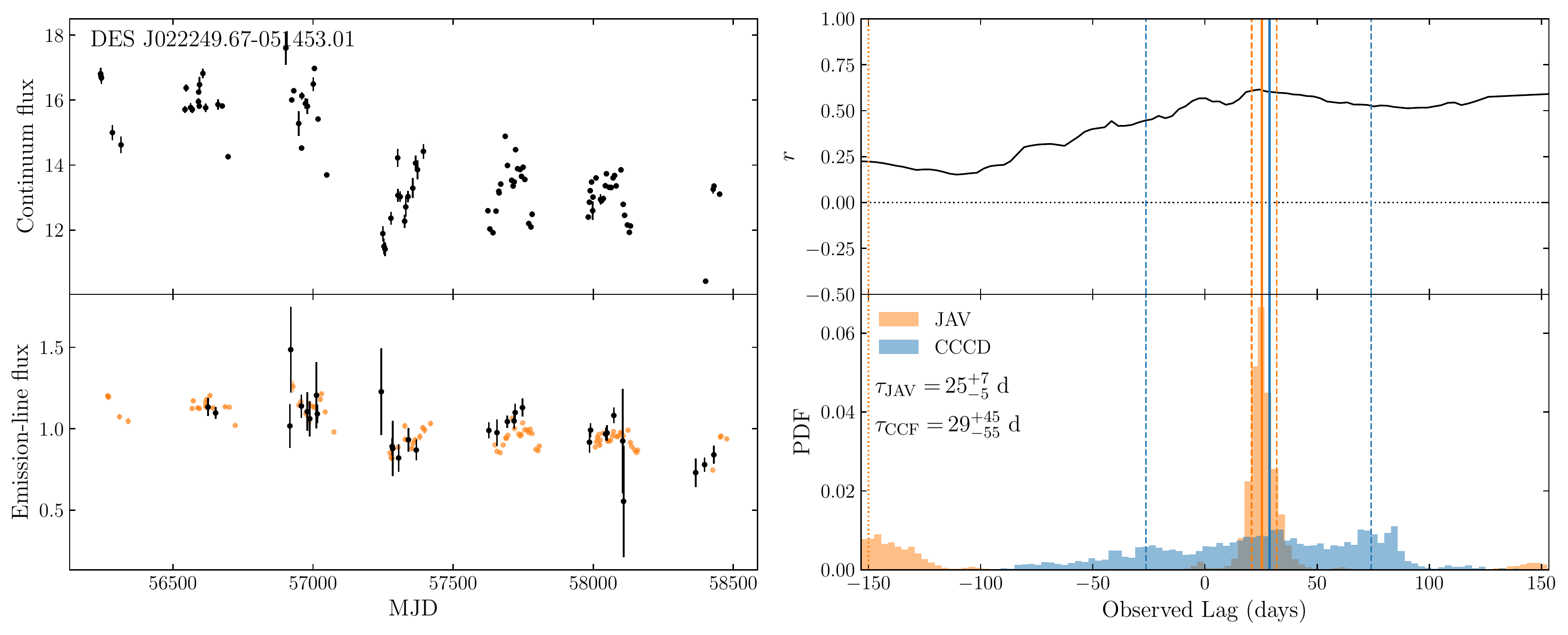}
    \end{subfigure}
    
    \begin{subfigure}[b]{\textwidth}
       \includegraphics[width=1\linewidth]{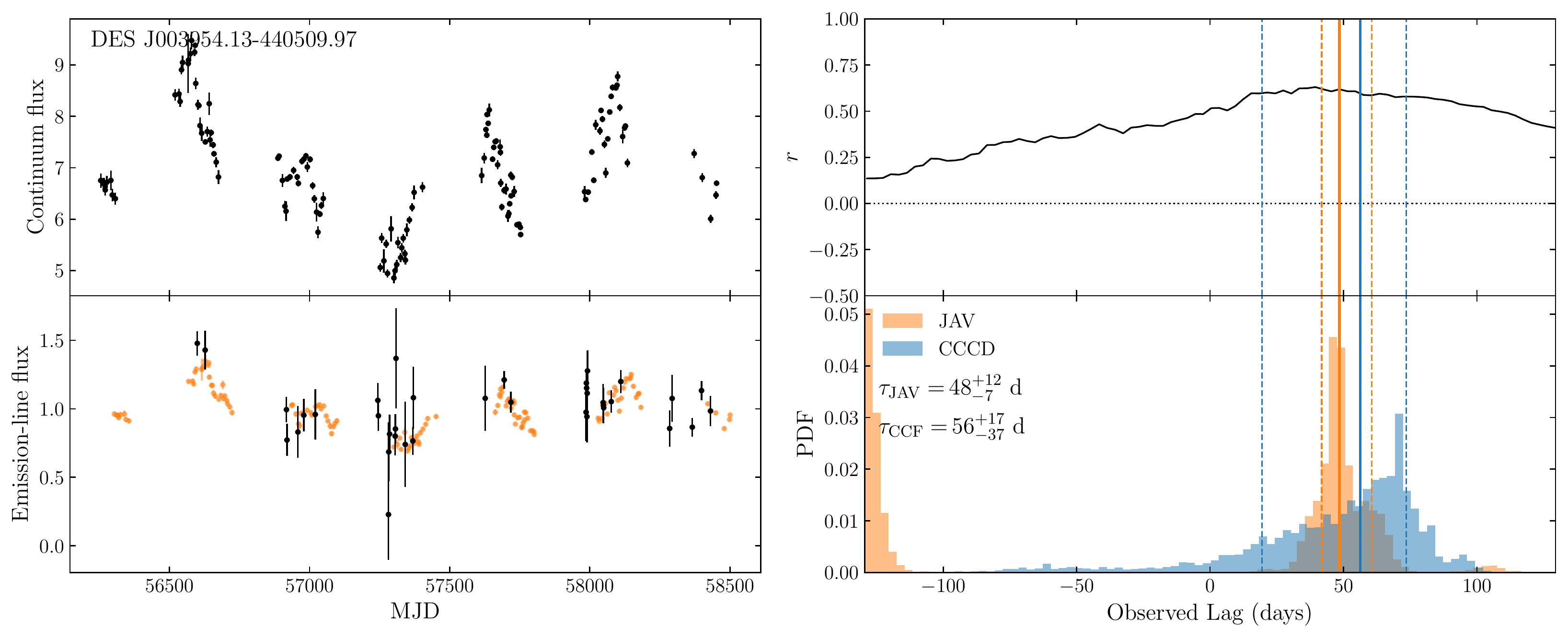}
    \end{subfigure}
        \caption{Same as for \autoref{fig:lags1}, for sources with \textsc{JAVELIN} aliasing signals. The orange vertical dotted lines indicate the negative aliasing peak in the \textsc{JAVELIN} PDF, for which we show the phase shifted light curves in Appendix \ref{sec:app_aliasing}. These three sources are included in the final sample. }
    \label{fig:aliasinglags1}
\end{figure*}

After visually inspecting the lag recoveries from our full H$\beta$ sample, we identified eight sources with lag PDF's of comparable quality to the sample that passed our selection criteria, but with \textsc{JAVELIN} aliasing peaks. This signal is produced as an artefact of the survey window function; predominantly the seasonal gaps. These aliasing peaks do not coincide with a peak in the \textsc{PyCCF} CCCD or CCF, and always occur at negative lags coincident with the seasonal gaps in the observational campaign ($\sim-150$ to $-250$ days). We show the light curves and lag PDF's of these sources in \autoref{fig:aliasinglags1}, \autoref{fig:aliasinglags2} and \autoref{fig:aliasinglags3}. For most sources, there is good agreement in the light curves phase-shifted by the positive recovered lag. For three of the eight sources (\autoref{fig:aliasinglags1}), there is overlap between the photometric and spectroscopic observations, while the light curves for the other five (\autoref{fig:aliasinglags2} and \autoref{fig:aliasinglags3}) shift such that the spectroscopic observations fall wholly within gaps in the photometric observations. For comparison, we phase-shifted the continuum light curve by the lag at which the negative \textsc{JAVELIN} peak occurs. We show the light curves phase-shifted by both the positive lag and the negative lag in \autoref{fig:aliasingLC1} and \autoref{fig:aliasingLC2}. In some cases, the light curves show smooth multi-year variations, for which the negatively phase-shifted light curves do seem to reasonably interpolate between the seasonal gaps; however, there is no coincident peak in the \textsc{PyCCF} CCCD.

If we omit the negative peaks in the \textsc{JAVELIN} PDF's, five of the eight sources pass our selection criteria, and therefore illustrate comparable quality to the sample recovered originally. Given there is no physical motivation for a negative reverberation lag, and the negative aliasing peaks seem to be an artefact of the \textsc{JAVELIN} method alone, we choose to include the three sources that demonstrate overlap between their light curves when shifted by the positive lag in our final sample. We provide the lags for these three sources in \autoref{tab:results}. At this stage we choose not to include the other two sources, DES J024533.65-000744.91 and DES J004009.06-431255.29, which have positive lags that cause the phase-shifted light curves to land completely within the seasonal gaps. The three sources that do not pass all our selection criteria only fail the $r_{\rm max}$ criterion, although we note that their $r_{\rm max}$ are above 0.5. The phase-shifted light curves for these three sources also fall within the seasonal gaps. High cadence monitoring to resolve shorter term variations, and observing over longer seasons will be required to reliably recover these lags. 

\autoref{fig:z_v_lum} shows the redshift and luminosity distribution of our final recovered sample of eight AGN, compared with existing measurements. Our sample probes higher redshifts, and spans 1 dex in luminosity.

\subsection{Black hole masses and the $R-L$ relation}

\begin{figure*}
    \centering
    \includegraphics[width=\textwidth]{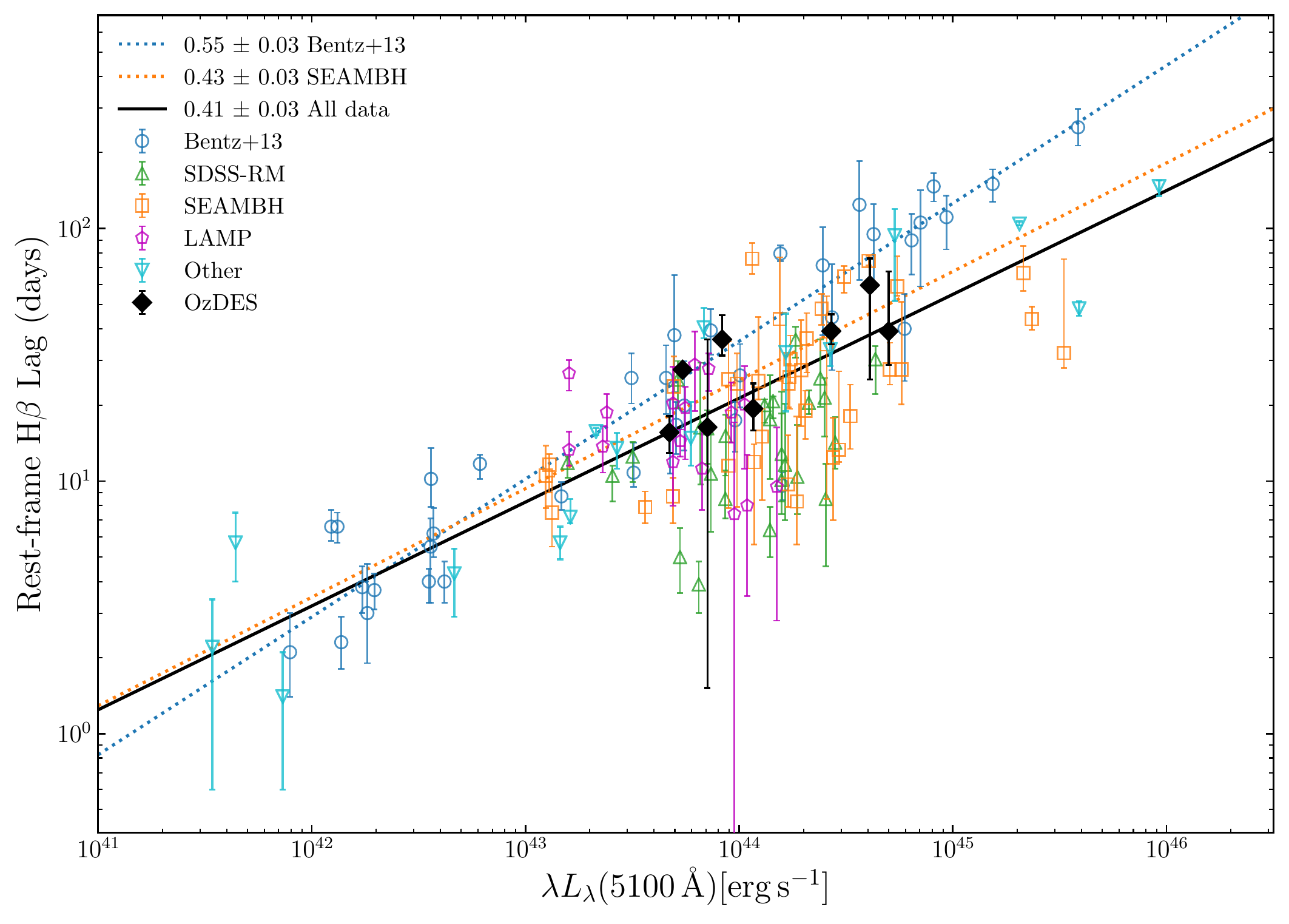}
    \caption{Radius-Luminosity relation for H$\beta$ using our final sample of eight new AGN together with existing measurements from \citet[][and references therein]{Bentz2013}; SDSS-RM \citep[][quality 4 and 5]{Grier2017}; SEAMBH  \citep{Du2014,Wang2014,Du2015,Du2016,Du2018,Hu2021}; Lick AGN Monitoring Project \citep[LAMP,][]{U2022}; and other measurements from \citet{Bentz2009,Barth2013,Bentz2014,Pei2014,Lu2016,Bentz2016a,Bentz2016b,Fausnaugh2017,Zhang2019,Rakshit2019,Li2021}, of which measurements published before 2019 are compiled by \citet{MartinezAldama2019}. The slope of the relation is given in the legend, along with the slopes constrained previously by \citet{Bentz2013} and \citet{Du2018}. All sources in the OzDES sample have low accretion rates, apart from our highest luminosity measurement (right-most black point), which has a moderately high accretion rate (\autoref{tab:results}). }
    \label{fig:RL}
\end{figure*}

We measured the H$\beta$ line-width using the line dispersion of the mean spectra. Although H$\beta$ line-width measurements are commonly made using the root mean squared (rms) spectra, the signal-to-noise of our spectra are insufficient to support this approach. We measure $M_{\rm BH}$ for our final sample of eight AGN using Equation \ref{virial} and a virial factor $f=4.47$ \citep{Woo2015}, and give these in \autoref{tab:results}, along with the line dispersion measurements. 

We constrain the $R-L$ relation as shown in \autoref{fig:RL}. Our best fit to the existing data and our final sample is 
\begin{equation}
    \log(R_{\rm BLR}/{\rm lt-day}) = K + \alpha \log(\lambda L_{\lambda})/10^{44} [{\rm erg\,s}^{-1}]
\end{equation}
with slope $\alpha=0.41\pm0.03$, $K=1.33\pm0.02$, and an intrinsic scatter of $\sigma=0.23\pm0.02$ dex. 

The SEAMBH program targeted highly accreting AGN at $z<0.4$ \citep{Du2014,Wang2014,Du2015,Du2016,Du2018,Hu2021}.
SEAMBH found that highly accreting AGN have systematically shorter reverberation lags. They proposed that accretion rate explains the observed deviation from the steeper $R-L$ relation from the \citet{Bentz2013} compilation of earlier results. Following \citet{Du2018}, we measure the dimensionless accretion rate of our sample using the estimator derived from the standard thin accretion disk model \citep{Shakura1973}:
\begin{equation}
    \dot{M} = 20.1 \left( \frac{l_{44}}{\cos{(i)}} \right) ^{3/2} m_{7}^{-2},
\end{equation}
where $l_{44} = L_{5100}/10^{44}$ erg\,s$^{-1}$, $m_7 = M_{\rm BH}/10^7$  M$_{\sun}$, and the inclination angle of the disk to the line of sight is taken to be $\cos(i) = 0.75$. All but one of our sources have low accretion rates, which is consistent with the observed agreement between our results and the \citet{Bentz2013} relation, which was constrained using mainly low-accretion sources. One of our sources is on the lower boundary of what SEAMBH define to be highly accreting, and is consistent with both the low-accretion and high-accretion samples.

\section{Discussion}\label{sec:discussion}

The primary goal of the OzDES RM Program was to make RM measurements at high redshifts, with the Mg\,\textsc{ii} and C\,\textsc{iv} lines. Our monthly cadence was sufficient for this, due to the intrinsically longer reverberation lags for these high ionisation lines, and the increased time dilation for such sources that are typically at higher redshifts than our H$\beta$ sample. Simulations showed the OzDES observational cadence was likely insufficient to provide a full sample of H$\beta$ RM measurements \citep{King2015,Malik2022}. Since our survey is not as sensitive to these shorter lags, some of our selection criteria are not as strict as other analyses. Some of the uncertainties on the lags we have recovered are considerably larger than previous works, as we lack the temporal resolution to recover lags precisely.

Our new measurements are more consistent with the \citet{Bentz2013} slope than the SDSS-RM H$\beta$ sample \citep{Grier2017}, although our sample has a similar redshift-luminosity distribution to SDSS-RM. As discussed by \citet{Grier2017}, the SDSS-RM lags may underestimate $M_{\rm BH}$ due to selection effects from having only a single campaign season of observation which limits the lag search window to 100d. Our OzDES analysis is based on a 7 years of photometry and 6 years of spectroscopy but with lower observational cadence. However, simulations by \citet{FonsecaAlvarez2020} suggest this deviation is not due to observational biases. They attribute the deviation of their sample to changes in the UV/optical spectral energy distribution (SED), although this was not measured directly. The only significant difference between the OzDES and SDSS-RM samples and lag analyses is the baseline and cadence of the light curve data.

Simulations presented by \citet{Malik2022} modelled survey window functions for OzDES and upcoming surveys, and investigated lag recovery efficacy with degraded or increased sampling. They find that with low sampling the scatter in the recovered $R-L$ relation is increased, however there is no systematic offset from the input slope \citep[see Fig. 13 in][]{Malik2022}. For sources with shorter expected lags and with a less-than-ideal cadence, they did not systematically recover longer lags. In the case that these idealised simulations are not representative of the data, it is possible that with our cadence we may not be as sensitive to shorter lags if they were present. Upcoming surveys, including the Time-Domain Extragalactic Survey (TiDES; \citealt{Swann2019}) and SDSS-V Black Hole Mapper \citep{Kollmeier2017}, which will be spectroscopically following up the Legacy Survey of Space and Time (LSST) deep-drilling fields, will be revisiting most of the OzDES fields with shorter spectroscopic cadence. These programs can investigate any potential discrepancies with our results due to the longer cadence of OzDES.

\section{Summary}\label{sec:summary}

We successfully recover reverberation lags with H$\beta$ for eight AGN from the six-year observations from DES and OzDES. These results from our multi-object survey are consistent with previous H$\beta$ analyses done with source-by-source observations, including the early results compiled in \citet{Bentz2013}. Our observations seem inconsistent with the large scatter observed in the SDSS-RM H$\beta$ sample \citep{Grier2017}, although the only significant difference between our analyses is the baseline and cadence of the survey data. Our sample includes only one moderately high accretion rate source, and the location of our final sample on the $R-L$ relation is consistent with earlier measurements with similar source accretion rates. This work compliments the higher redshift results for the rest of the OzDES RM sample of 735 AGN, made with the Mg\,\textsc{ii} \citep[][]{Yu2021,Yu2022} and C\,\textsc{iv} lines \citep[][Penton et al. in prep]{Hoormann2019,Penton2022}, which will be used to recalibrate black hole mass-scaling relations for those emission lines.

The H$\beta$ sample is highly sensitive to the observational window function. Additional campaign seasons seem to help, but higher cadence and observing over longer seasons is necessary to reliably recover these lags. Future surveys, including TiDES \citep{Swann2019} and SDSS-V BHM \citep{Kollmeier2017}, which will be following up LSST, will observe some of the same fields as OzDES. These surveys can follow up with a more suitable window function \citep{Malik2022}. In order to anchor the ends of the H$\beta$ $R-L$ relation, future programs will need to target lower luminosity sources at $10^{41} < \lambda L_{\lambda}\,(5100$\,\AA) $< 10^{43}$ ergs\,s$^{-1}$, and higher luminosity sources at $\lambda L_{\lambda}\,(5100$\,\AA) $>10^{45}$ ergs\,s$^{-1}$. The challenge is to have a survey area large enough to observe local sources at low luminosity, and the rare local high luminosity sources. This is especially difficult for multi-object RM surveys, but by extending spectral coverage into the red-optical and infra-red range, such surveys can access the greater volume for such sources at higher redshifts with H$\beta$.

\section*{Acknowledgements}

We thank the anonymous referee for their comments that improved the paper. UM and AP are supported by the Australian Government Research Training Program (RTP) Scholarship. PM and ZY were supported in part by the United States National Science Foundation under Grant No. 161553 to PM. PM is grateful for support from the Radcliffe Institute for Advanced Study at Harvard University. PM also acknowledges support from the United States Department of Energy, Office of High Energy Physics under Award Number DE-SC-0011726. TMD is supported by an Australian Research Council Laureate Fellowship (project number FL180100168). 

We acknowledge parts of this research were carried out on the traditional lands of the Ngunnawal and Ngambri peoples. This work makes use of data acquired at the Anglo-Australian
Telescope, under program A/2013B/012. We acknowledge the Gamilaraay people as the traditional owners of the land on which the AAT stands. We pay our respects to their elders past, present, and emerging.

This analysis used \textsc{NumPy} \citep{harris2020array}, \textsc{Astropy} \citep{astropy:2013, astropy:2018}, and \textsc{SciPy} \citep{2020SciPy-NMeth}. Plots were made using \textsc{Matplotlib} \citep{Hunter:2007}. This work has made use of the SAO/NASA Astrophysics Data System Bibliographic Services.

This paper has gone through internal review by the DES collaboration. Funding for the DES Projects has been provided by the U.S. Department of Energy, the U.S. National Science Foundation, the Ministry of Science and Education of Spain, 
the Science and Technology Facilities Council of the United Kingdom, the Higher Education Funding Council for England, the National Center for Supercomputing 
Applications at the University of Illinois at Urbana-Champaign, the Kavli Institute of Cosmological Physics at the University of Chicago, 
the Center for Cosmology and Astro-Particle Physics at the Ohio State University,
the Mitchell Institute for Fundamental Physics and Astronomy at Texas A\&M University, Financiadora de Estudos e Projetos, 
Funda{\c c}{\~a}o Carlos Chagas Filho de Amparo {\`a} Pesquisa do Estado do Rio de Janeiro, Conselho Nacional de Desenvolvimento Cient{\'i}fico e Tecnol{\'o}gico and 
the Minist{\'e}rio da Ci{\^e}ncia, Tecnologia e Inova{\c c}{\~a}o, the Deutsche Forschungsgemeinschaft and the Collaborating Institutions in the Dark Energy Survey. 

The Collaborating Institutions are Argonne National Laboratory, the University of California at Santa Cruz, the University of Cambridge, Centro de Investigaciones Energ{\'e}ticas, 
Medioambientales y Tecnol{\'o}gicas-Madrid, the University of Chicago, University College London, the DES-Brazil Consortium, the University of Edinburgh, 
the Eidgen{\"o}ssische Technische Hochschule (ETH) Z{\"u}rich, 
Fermi National Accelerator Laboratory, the University of Illinois at Urbana-Champaign, the Institut de Ci{\`e}ncies de l'Espai (IEEC/CSIC), 
the Institut de F{\'i}sica d'Altes Energies, Lawrence Berkeley National Laboratory, the Ludwig-Maximilians Universit{\"a}t M{\"u}nchen and the associated Excellence Cluster Universe, 
the University of Michigan, NSF's NOIRLab, the University of Nottingham, The Ohio State University, the University of Pennsylvania, the University of Portsmouth, 
SLAC National Accelerator Laboratory, Stanford University, the University of Sussex, Texas A\&M University, and the OzDES Membership Consortium.

Based in part on observations at Cerro Tololo Inter-American Observatory at NSF's NOIRLab (NOIRLab Prop. ID 2012B-0001; PI: J. Frieman), which is managed by the Association of Universities for Research in Astronomy (AURA) under a cooperative agreement with the National Science Foundation.

The DES data management system is supported by the National Science Foundation under Grant Numbers AST-1138766 and AST-1536171.
The DES participants from Spanish institutions are partially supported by MICINN under grants ESP2017-89838, PGC2018-094773, PGC2018-102021, SEV-2016-0588, SEV-2016-0597, and MDM-2015-0509, some of which include ERDF funds from the European Union. IFAE is partially funded by the CERCA program of the Generalitat de Catalunya.
Research leading to these results has received funding from the European Research
Council under the European Union's Seventh Framework Program (FP7/2007-2013) including ERC grant agreements 240672, 291329, and 306478.
We  acknowledge support from the Brazilian Instituto Nacional de Ci\^encia
e Tecnologia (INCT) do e-Universo (CNPq grant 465376/2014-2).

This manuscript has been authored by Fermi Research Alliance, LLC under Contract No. DE-AC02-07CH11359 with the U.S. Department of Energy, Office of Science, Office of High Energy Physics.

%%%%%%%%%%%%%%%%%%%%%%%%%%%%%%%%%%%%%%%%%%%%%%%%%%
\section*{Data Availability}
Machine-readable light curves for the eight AGN in the final sample are available in the online supplementary material. The underlying DES and OzDES data are available in \citet{Abbott2021} and \citet{Lidman2020}. Please note: Oxford University Press is not responsible for the content or functionality of any supporting materials supplied by the authors. Any queries (other than missing material) should be directed to the corresponding author for the article.
 
% The inclusion of a Data Availability Statement is a requirement for articles published in MNRAS. Data Availability Statements provide a standardised format for readers to understand the availability of data underlying the research results described in the article. The statement may refer to original data generated in the course of the study or to third-party data analysed in the article. The statement should describe and provide means of access, where possible, by linking to the data or providing the required accession numbers for the relevant databases or DOIs.

%%%%%%%%%%%%%%%%%%%%%%%%%%%%%%%%%%%%%%%%%%%%%%%%%%

%%%%%%%%%%%%%%%%%%%% REFERENCES %%%%%%%%%%%%%%%%%%

% The best way to enter references is to use BibTeX:

\bibliographystyle{mnras}
\bibliography{refs} % if your bibtex file is called example.bib

%%%%%%%%%%%%%%%%%%%%%%%%%%%%%%%%%%%%%%%%%%%%%%%%%%

%%%%%%%%%%%%%%%%% APPENDICES %%%%%%%%%%%%%%%%%%%%%

\appendix

\section{Sources with \textsc{JAVELIN} aliasing} \label{sec:app_aliasing}

\begin{figure*}
    \captionsetup[subfigure]{slc=off}
    \centering
    \begin{subfigure}[b]{\textwidth}
       \includegraphics[width=1\linewidth]{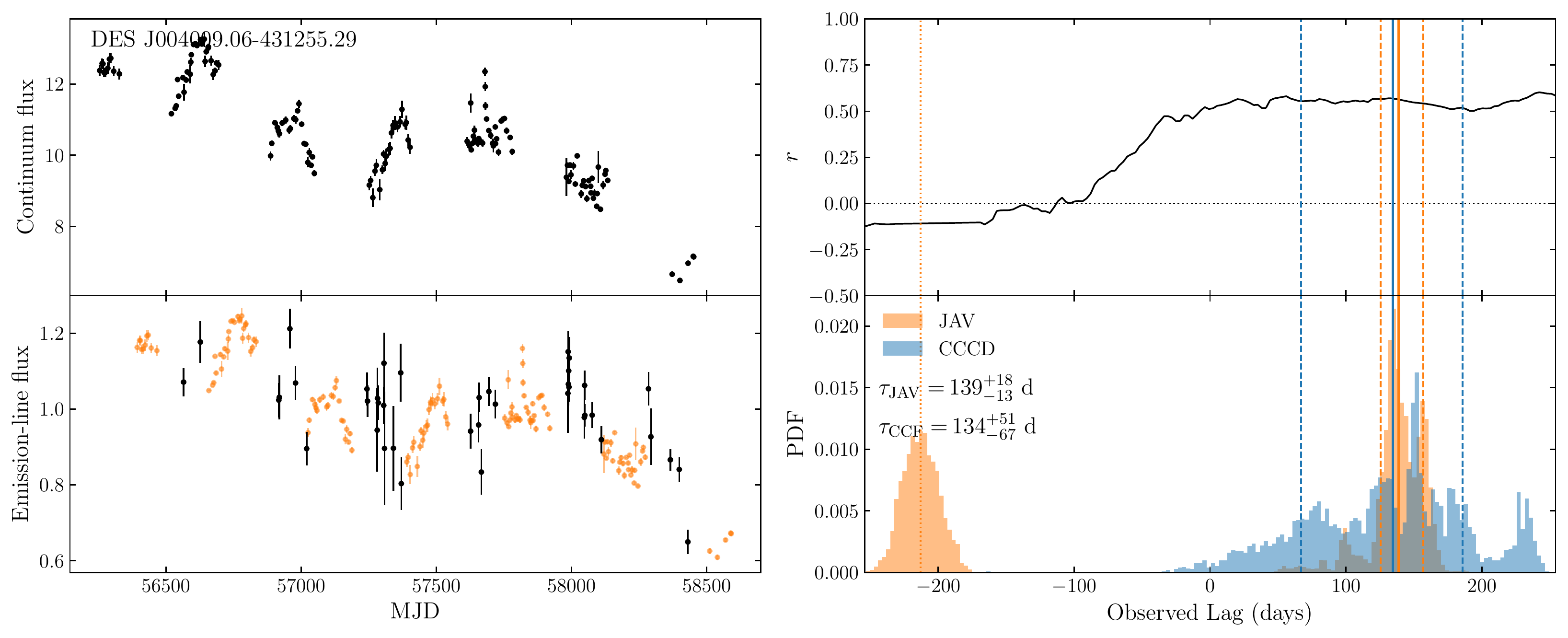}
    \end{subfigure}
    \begin{subfigure}[b]{\textwidth}
       \includegraphics[width=1\linewidth]{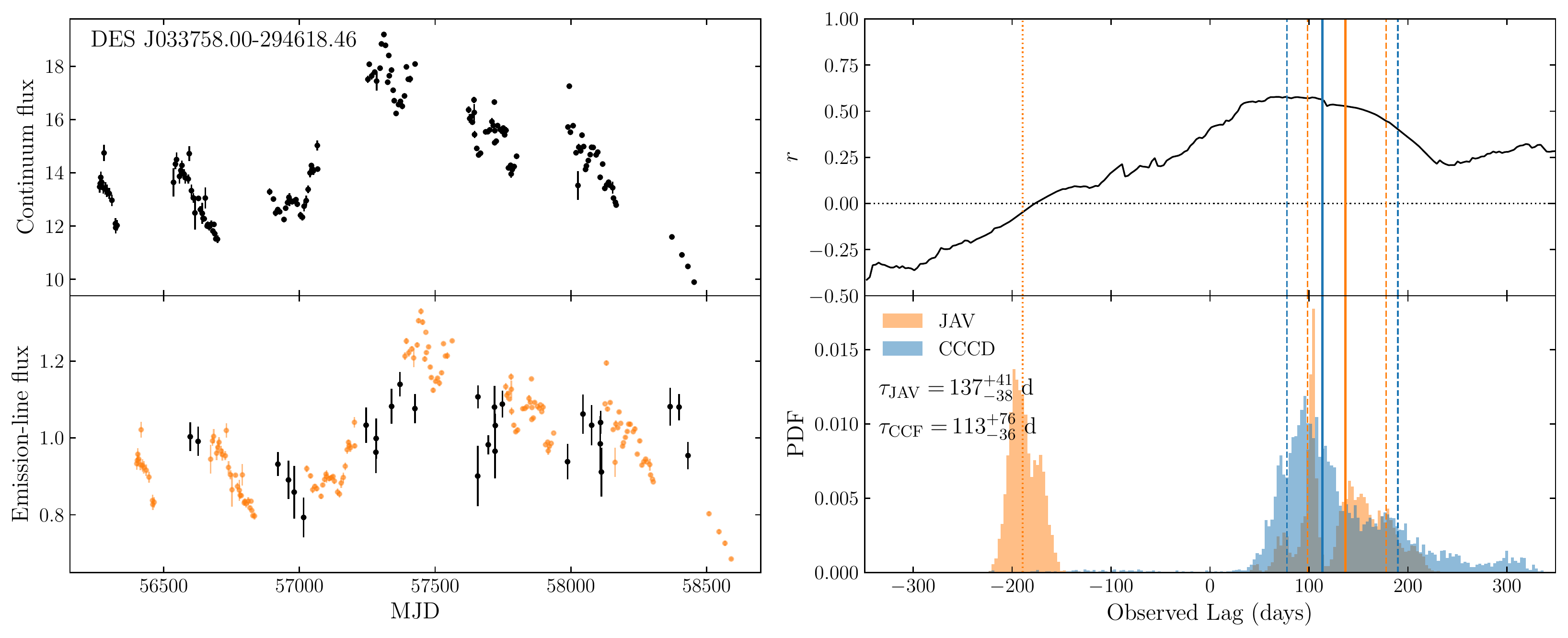}
    \end{subfigure}
    \begin{subfigure}[b]{\textwidth}
       \includegraphics[width=1\linewidth]{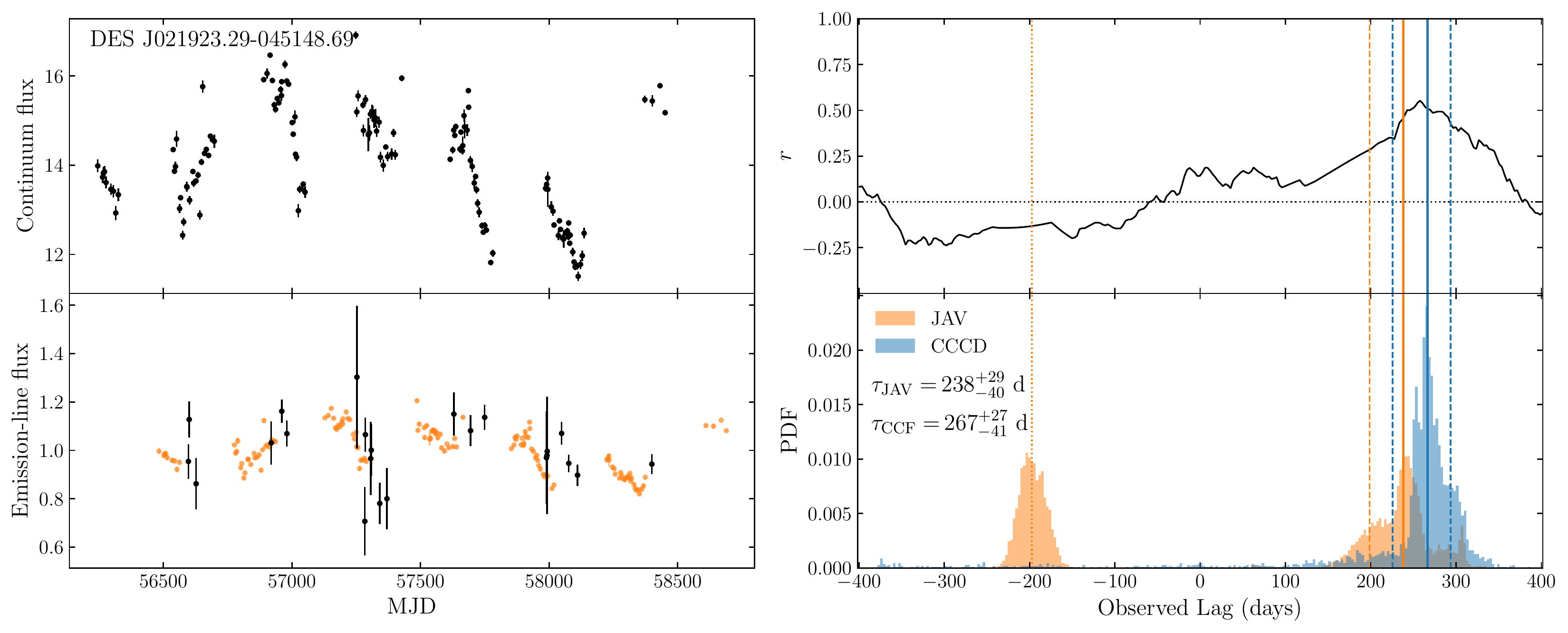}
    \end{subfigure}
    \caption{Same as for \autoref{fig:aliasinglags1}. These sources are not included in the final sample.}
    \label{fig:aliasinglags2}
\end{figure*}

\begin{figure*}
    \captionsetup[subfigure]{slc=off}
    \centering
    \begin{subfigure}[b]{\textwidth}
       \includegraphics[width=1\linewidth]{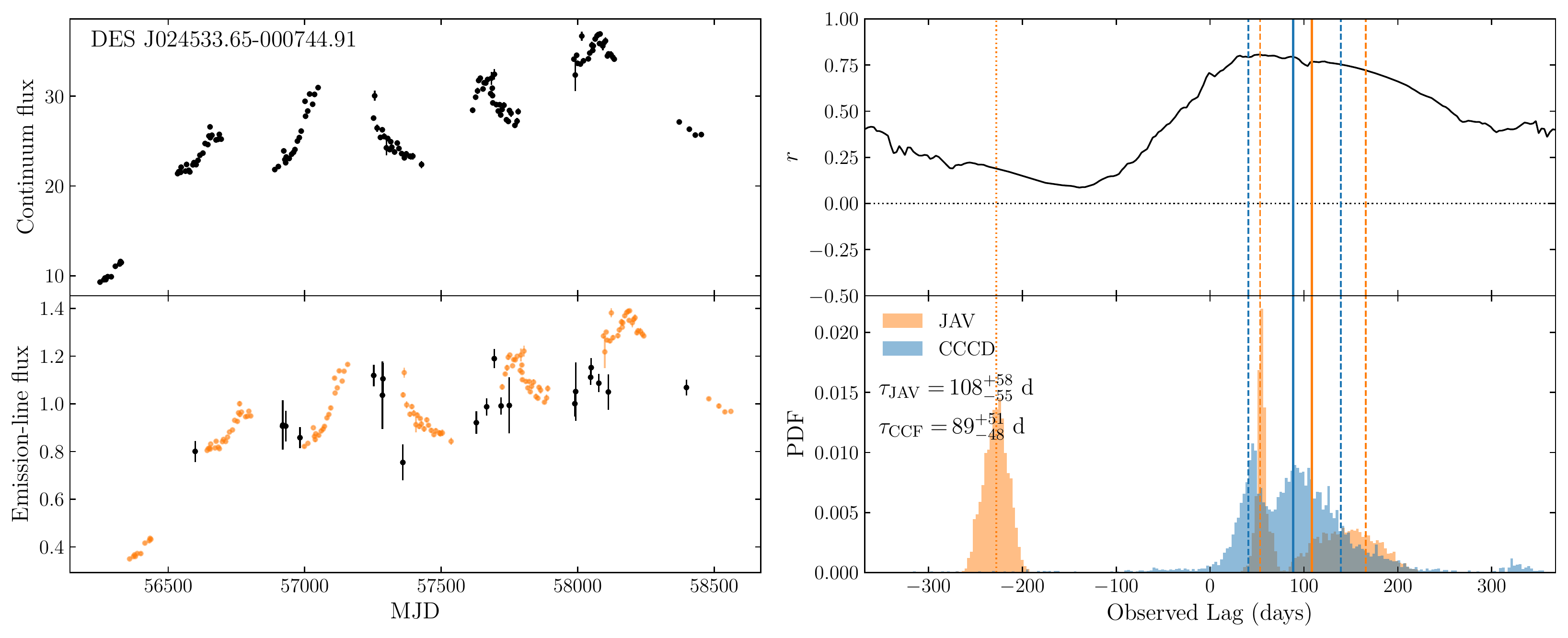}
    \end{subfigure}
    \begin{subfigure}[b]{\textwidth}
       \includegraphics[width=1\linewidth]{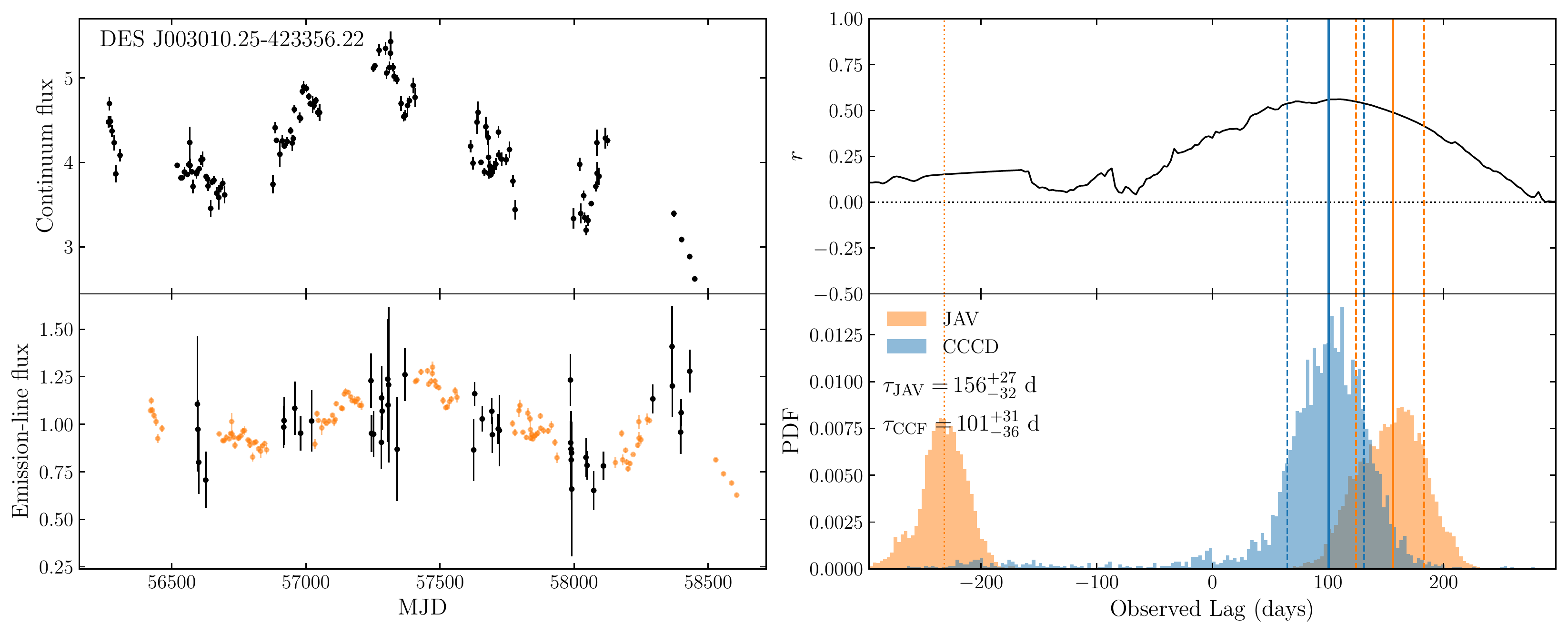}
    \end{subfigure}
    \caption{Same as for \autoref{fig:aliasinglags1}. These sources are not included in the final sample.}
    \label{fig:aliasinglags3}
\end{figure*}

\begin{figure}
    \captionsetup[subfigure]{slc=off}
    \centering
    \begin{subfigure}[b]{0.45\textwidth}
       \includegraphics[width=\textwidth]{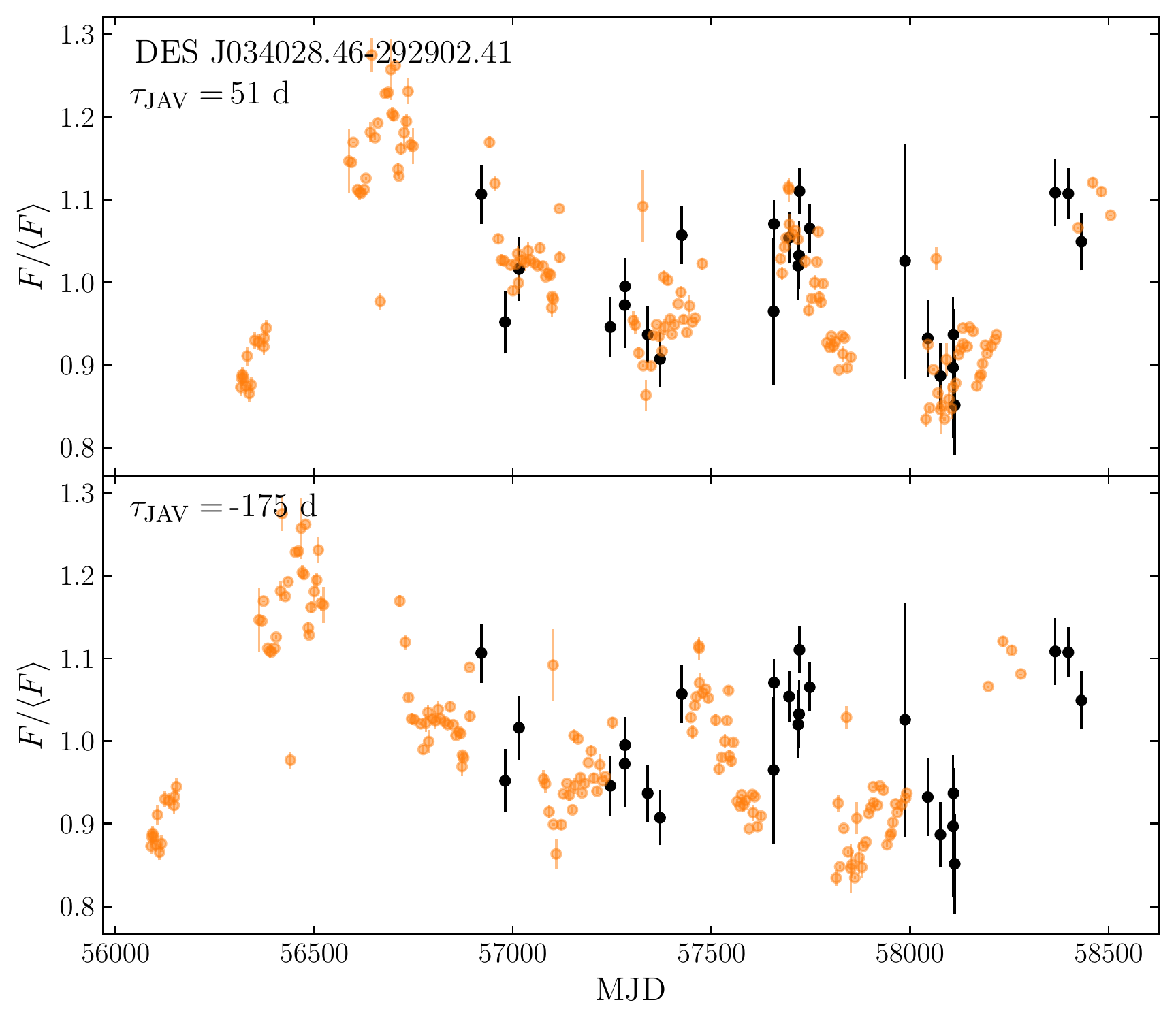}
    \end{subfigure}
    
    \begin{subfigure}[b]{0.45\textwidth}
       \includegraphics[width=\textwidth]{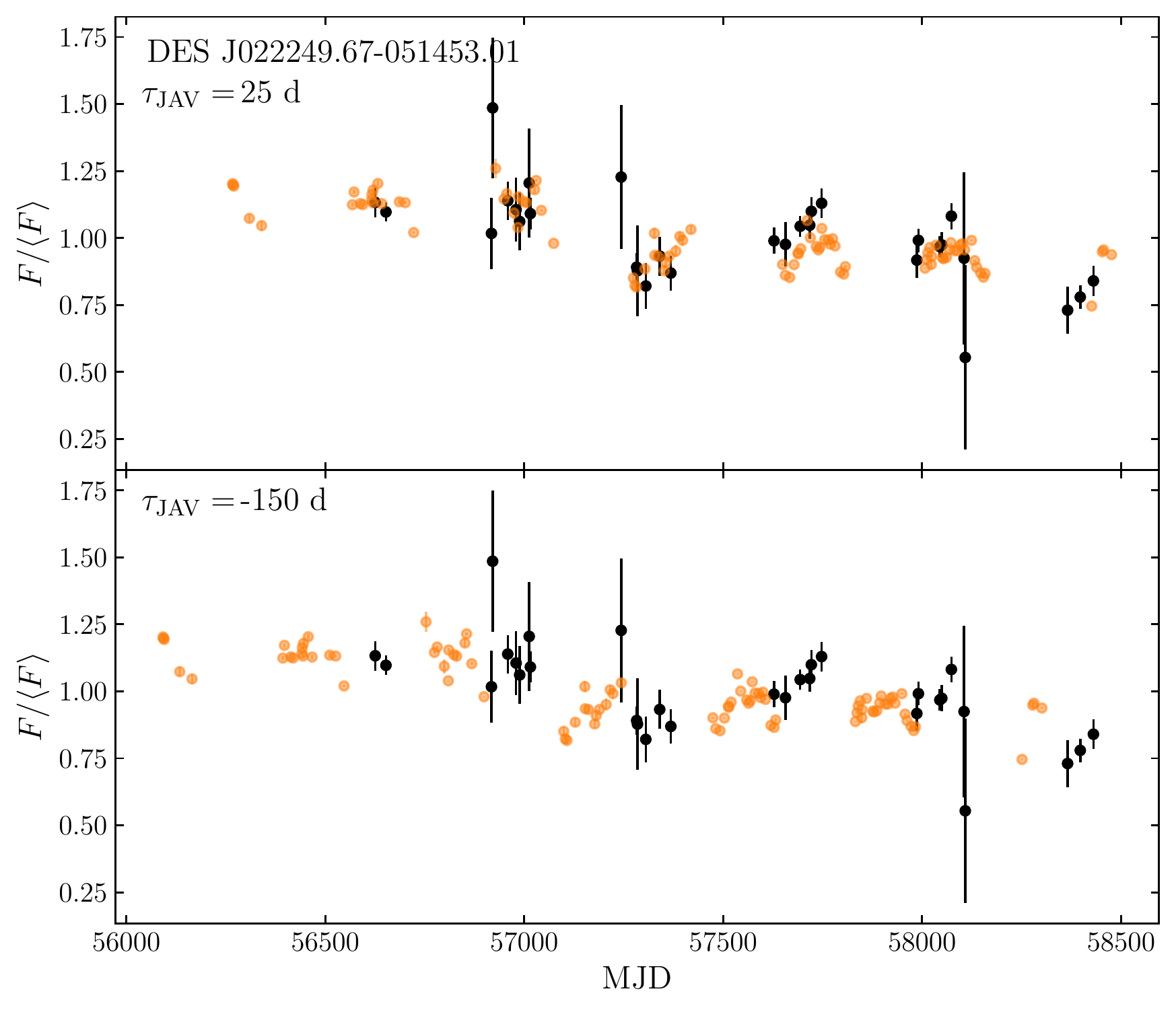}
    \end{subfigure}
    
    \begin{subfigure}[b]{0.45\textwidth}
       \includegraphics[width=\textwidth]{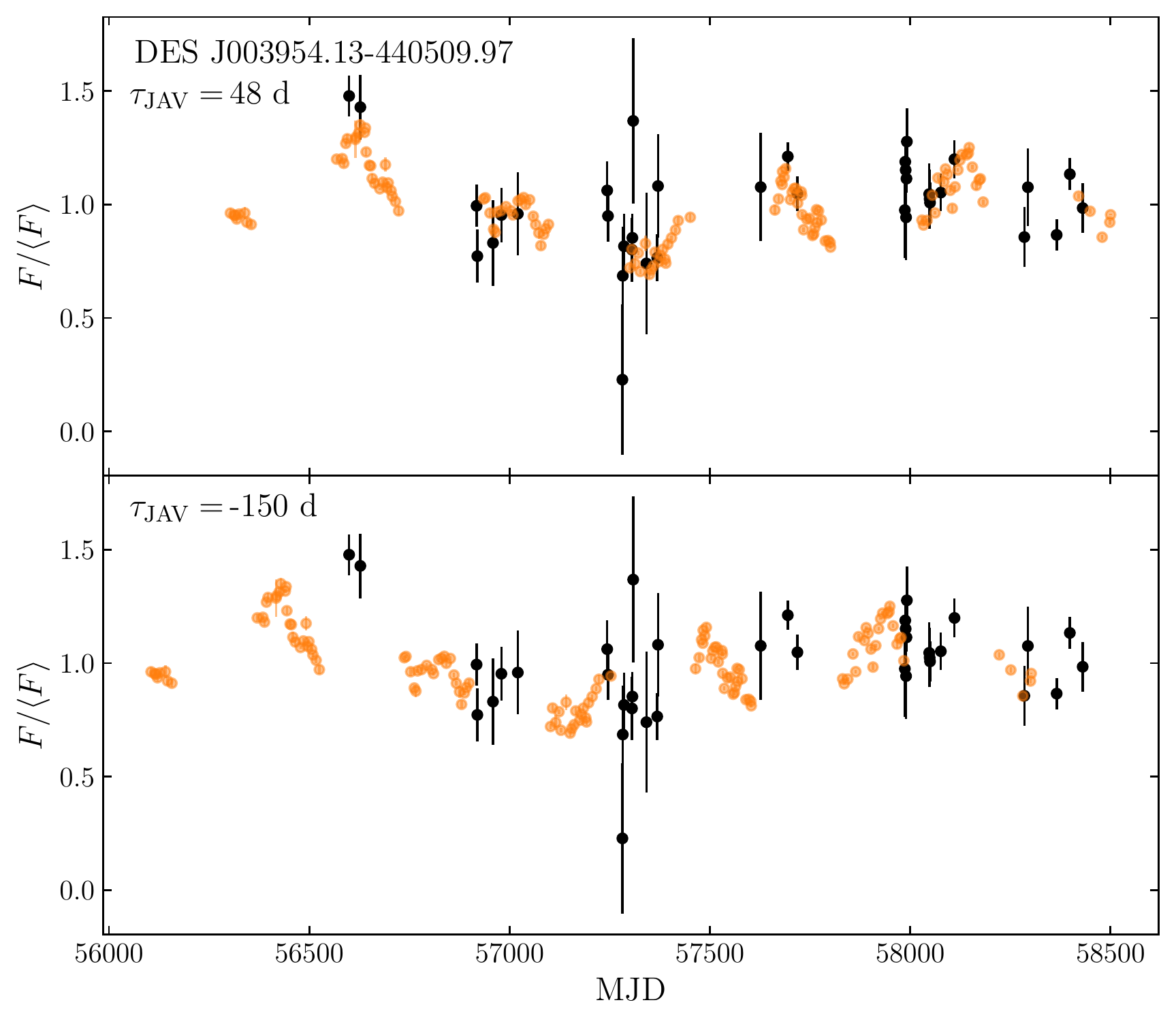}
    \end{subfigure}
        \caption{The emission-line light curve (black) and phase-shifted continuum light curve (orange) for the additional three sources we include in our final sample. The top and bottom panels show the continuum shifted by the positive and negative lag, respectively. For each source there is overlap between the light curves phase-shifted by the positive lag, while the negative lag shifts the continuum light curves wholly within the seasonal gaps of the emission-line light curve. }
    \label{fig:aliasingLC1}
\end{figure}

\begin{figure*}
    \centering
    \begin{subfigure}[b]{0.45\textwidth}
        \includegraphics[width=\textwidth]{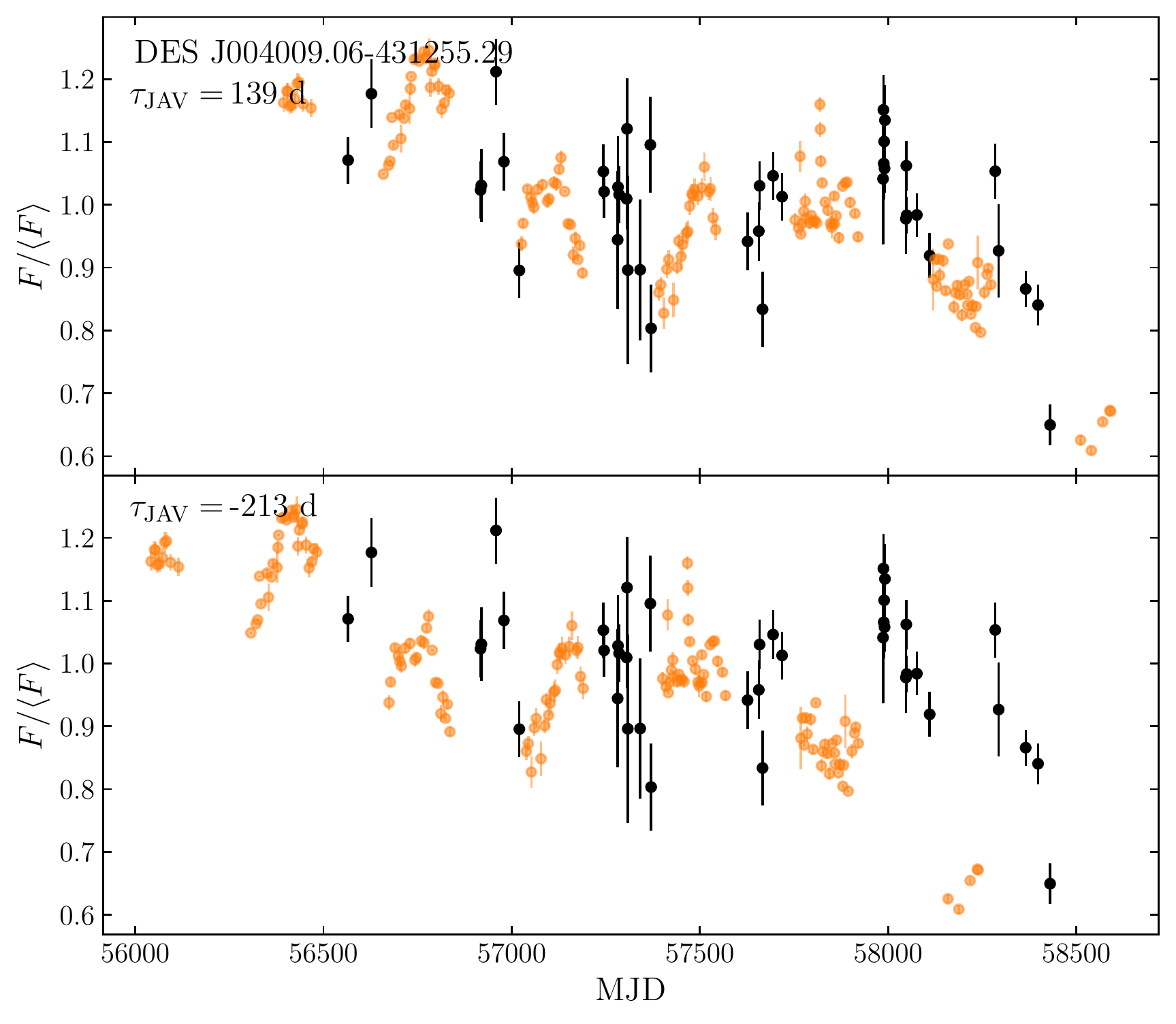}
    \end{subfigure}
    \medskip
    \begin{subfigure}[b]{0.45\textwidth}
       \includegraphics[width=\textwidth]{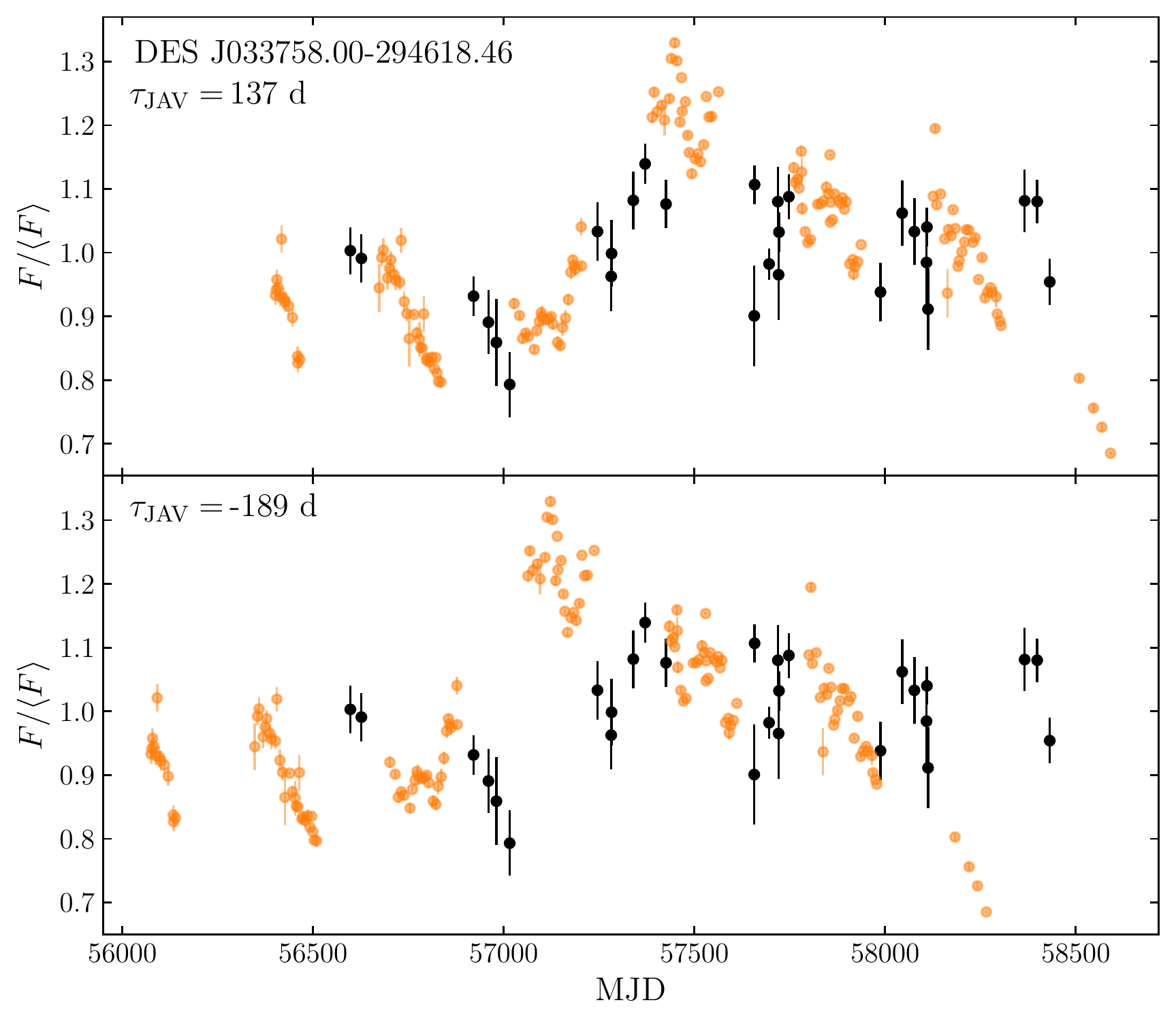}
    \end{subfigure}
    
    \begin{subfigure}[b]{0.45\textwidth}  
        \includegraphics[width=\textwidth]{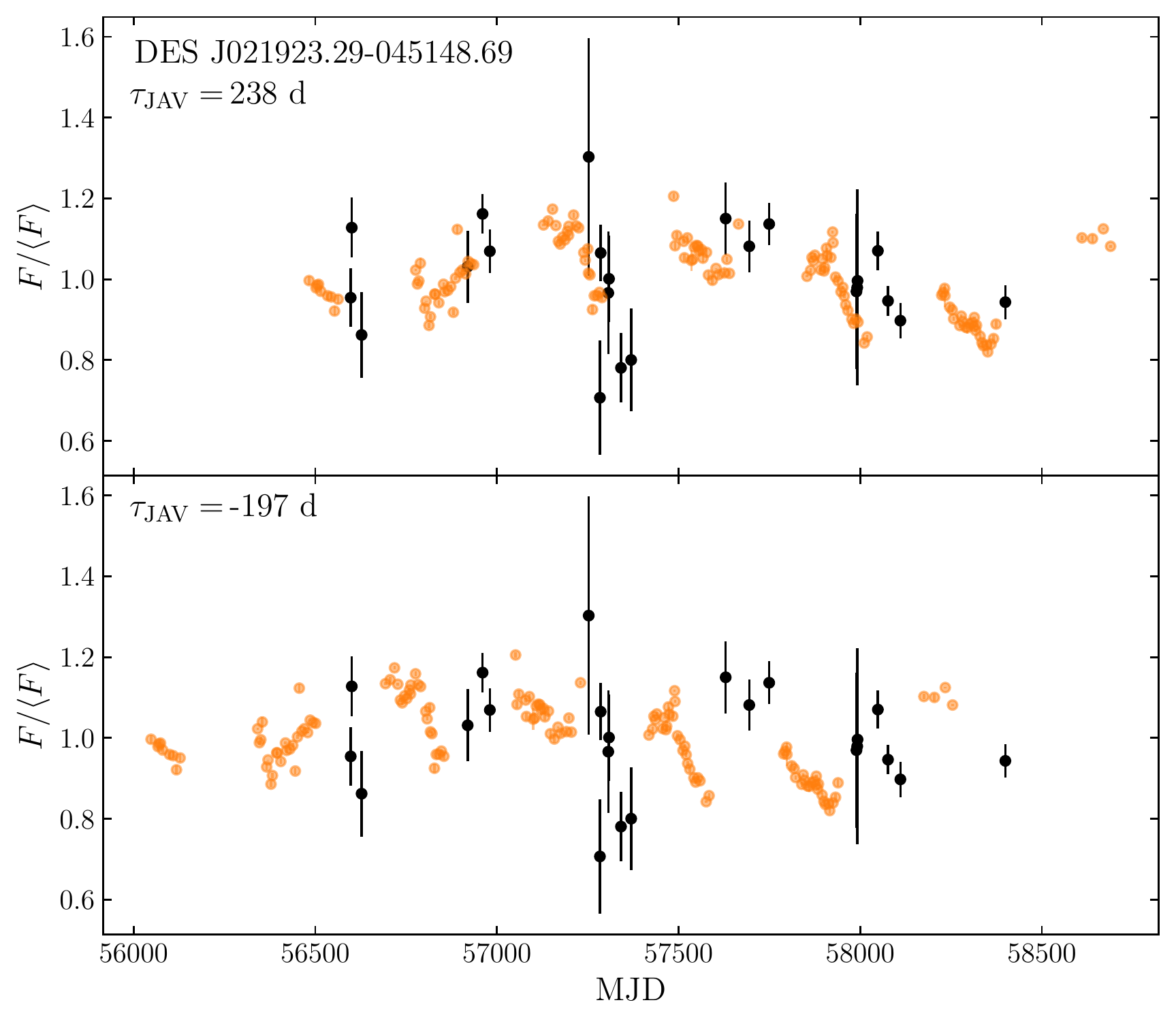}
    \end{subfigure}
    \medskip
    \begin{subfigure}[b]{0.45\textwidth}   
        \includegraphics[width=\textwidth]{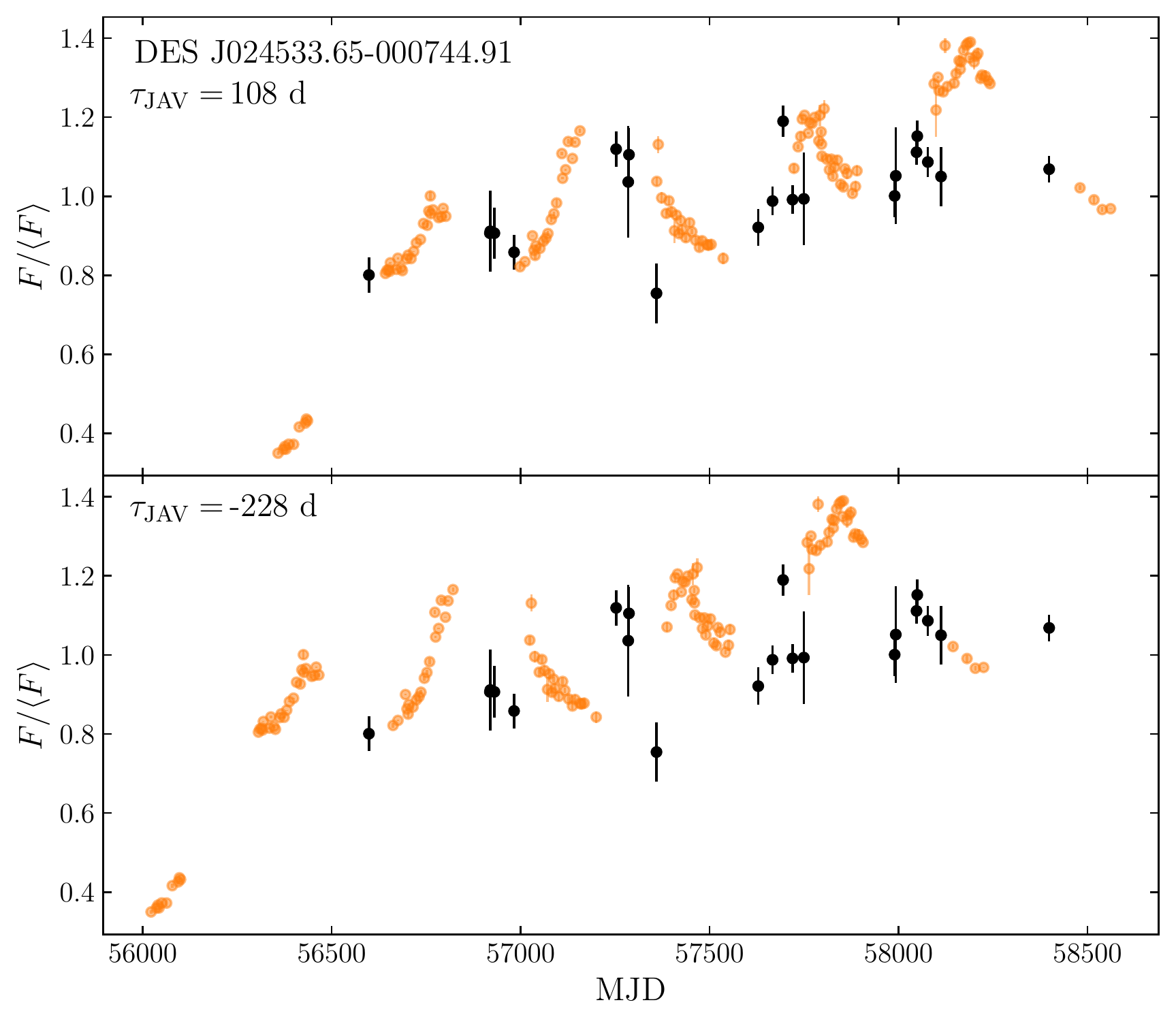}
    \end{subfigure}

    \begin{subfigure}[b]{0.45\textwidth}   
        \includegraphics[width=\textwidth]{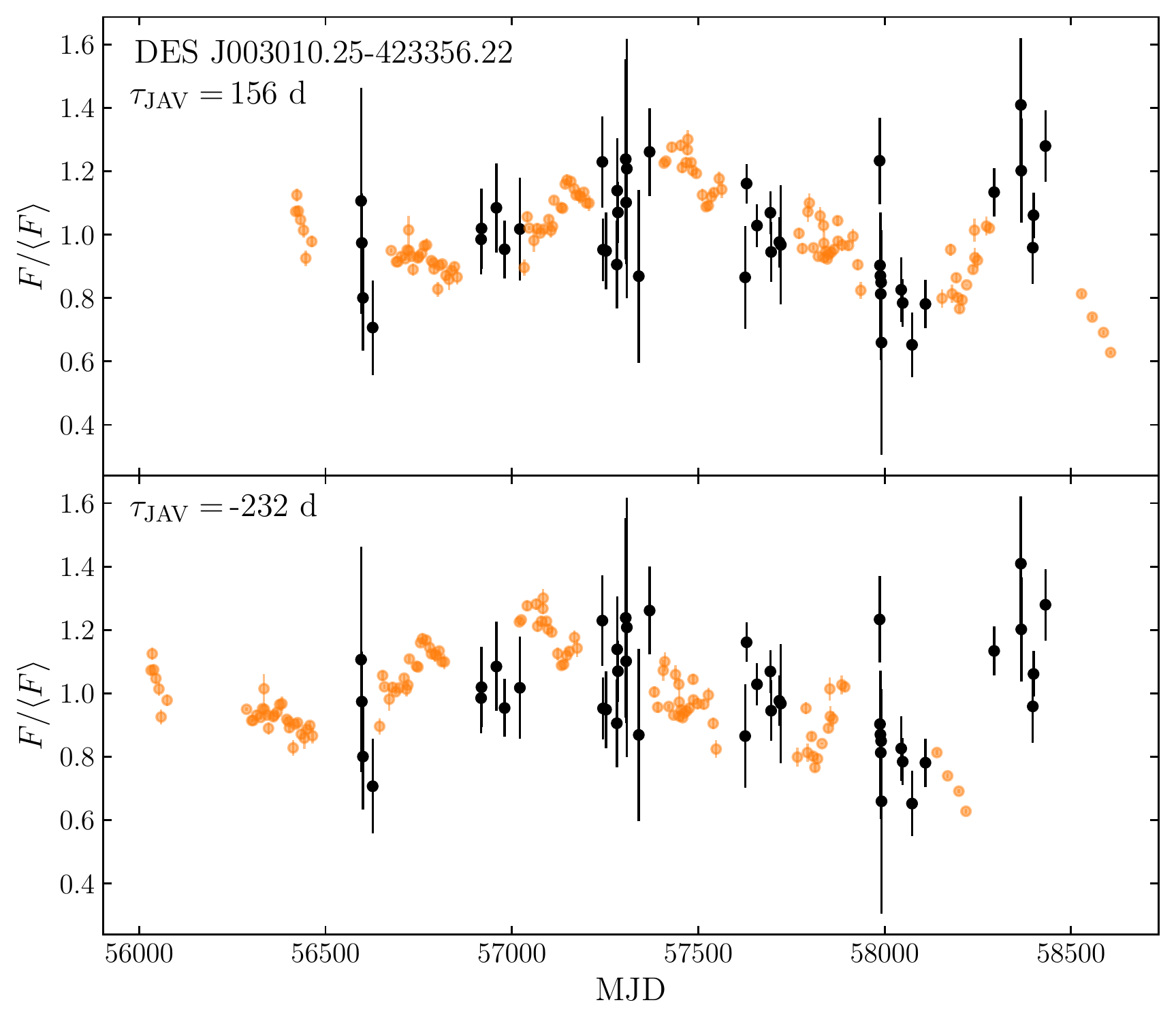}
    \end{subfigure}
    \caption[]{The emission-line light curve (black) and phase-shifted continuum light curve (orange) for the five sources in \autoref{fig:aliasinglags2} and \autoref{fig:aliasinglags3}, which we do not include in our final sample. The top and bottom panels show the continuum shifted by the positive and negative lag, respectively. For each source, in both cases there is no overlap between the photometric and spectroscopic observations.} 
    \label{fig:aliasingLC2}
\end{figure*}

Here we include additional figures referenced in \S\ref{sec:jav_aliasing}. \autoref{fig:aliasinglags2} and \autoref{fig:aliasinglags3} show the light curves and lag PDF's for the five of the eight sources with \textsc{JAVELIN} aliasing signals that we do not include on our final sample. Although two of these sources do pass our selection criteria after omitting the negative \textsc{JAVELIN} aliasing peak, we choose to not include them in our final sample as the positive lag is not constrained by overlap between the photometric and spectroscopic data. We show the light curves phase-shifted by both the positive lag and the negative lag in \autoref{fig:aliasingLC1} for the three sources which are included in our final sample, and in \autoref{fig:aliasingLC2} for the other five sources.

% If you want to present additional material which would interrupt the flow of the main paper,
% it can be placed in an Appendix which appears after the list of references.

% \vspace{0.4cm}
% \noindent \textbf{Affiliations}\\
\section*{Affiliations}
$^{1}$Research School of Astronomy and Astrophysics, Australian National University, Canberra, ACT 2611, Australia\\
$^{2}$School of Mathematics and Physics, The University of Queensland,  St Lucia, QLD 4101, Australia\\
$^{3}$Department of Astronomy, The Ohio State University, Columbus, Ohio 43210, USA\\
$^{4}$Center of Cosmology and Astro-Particle Physics, The Ohio State University, Columbus, Ohio 43210, USA\\
$^{5}$Radcliffe Institute for Advanced Study, Harvard University, Cambridge, MA 02138, USA\\
$^{6}$National Centre for the Public Awareness of Science, Australian National University, Canberra, ACT 2601, Australia\\
$^{7}$The Australian Research Council Centre of Excellence for All-Sky Astrophysics in 3 Dimension (ASTRO 3D), Australia\\
$^{8}$Sydney Institute for Astronomy, School of Physics, A28, The University of Sydney, Sydney, NSW 2006, Australia
$^{9}$Laborat\'orio Interinstitucional de e-Astronomia - LIneA, Rua Gal. Jos\'e Cristino 77, Rio de Janeiro, RJ - 20921-400, Brazil\\
$^{10}$Fermi National Accelerator Laboratory, P. O. Box 500, Batavia, IL 60510, USA\\
$^{11}$Department of Physics, University of Michigan, Ann Arbor, MI 48109, USA\\
$^{12}$Centro de Investigaciones Energ\'eticas, Medioambientales y Tecnol\'ogicas (CIEMAT), Madrid, Spain\\
$^{13}$Institute of Cosmology and Gravitation, University of Portsmouth, Portsmouth, PO1 3FX, UK\\
$^{14}$CNRS, UMR 7095, Institut d'Astrophysique de Paris, F-75014, Paris, France\\
$^{15}$Sorbonne Universit\'es, UPMC Univ Paris 06, UMR 7095, Institut d'Astrophysique de Paris, F-75014, Paris, France\\
$^{16}$University Observatory, Faculty of Physics, Ludwig-Maximilians-Universit\"at, Scheinerstr. 1, 81679 Munich, Germany
$^{17}$Department of Physics \& Astronomy, University College London, Gower Street, London, WC1E 6BT, UK\\
$^{18}$Kavli Institute for Particle Astrophysics \& Cosmology, P. O. Box 2450, Stanford University, Stanford, CA 94305, USA\\
$^{19}$SLAC National Accelerator Laboratory, Menlo Park, CA 94025, USA\\
$^{20}$Instituto de Astrofisica de Canarias, E-38205 La Laguna, Tenerife, Spain\\
$^{21}$Universidad de La Laguna, Dpto. Astrofísica, E-38206 La Laguna, Tenerife, Spain\\
$^{22}$INAF-Osservatorio Astronomico di Trieste, via G. B. Tiepolo 11, I-34143 Trieste, Italy\\
$^{23}$Department of Astronomy, University of Illinois at Urbana-Champaign, 1002 W. Green Street, Urbana, IL 61801, USA\\
$^{24}$Center for Astrophysical Surveys, National Center for Supercomputing Applications, 1205 West Clark St., Urbana, IL 61801, USA\\
$^{25}$Institut de F\'{\i}sica d'Altes Energies (IFAE), The Barcelona Institute of Science and Technology, Campus UAB, 08193 Bellaterra (Barcelona) Spain\\
$^{26}$Institute for Fundamental Physics of the Universe, Via Beirut 2, 34014 Trieste, Italy\\
$^{27}$Astronomy Unit, Department of Physics, University of Trieste, via Tiepolo 11, I-34131 Trieste, Italy\\
$^{28}$Hamburger Sternwarte, Universit\"{a}t Hamburg, Gojenbergsweg 112, 21029 Hamburg, Germany\\
$^{29}$Department of Physics, IIT Hyderabad, Kandi, Telangana 502285, India\\
$^{30}$Jet Propulsion Laboratory, California Institute of Technology, 4800 Oak Grove Dr., Pasadena, CA 91109, USA\\
$^{31}$Institute of Theoretical Astrophysics, University of Oslo. P.O. Box 1029 Blindern, NO-0315 Oslo, Norway\\
$^{32}$Kavli Institute for Cosmological Physics, University of Chicago, Chicago, IL 60637, USA\\
$^{33}$Instituto de Fisica Teorica UAM/CSIC, Universidad Autonoma de Madrid, 28049 Madrid, Spain\\
$^{34}$Department of Astronomy, University of Michigan, Ann Arbor, MI 48109, USA\\
$^{35}$Observat\'orio Nacional, Rua Gal. Jos\'e Cristino 77, Rio de Janeiro, RJ - 20921-400, Brazil\\
$^{36}$Santa Cruz Institute for Particle Physics, Santa Cruz, CA 95064, USA\\
$^{37}$Center for Astrophysics $\vert$ Harvard \& Smithsonian, 60 Garden Street, Cambridge, MA 02138, USA\\
$^{38}$Australian Astronomical Optics, Macquarie University, North Ryde, NSW 2113, Australia\\
$^{39}$Lowell Observatory, 1400 Mars Hill Rd, Flagstaff, AZ 86001, USA\\
$^{40}$George P. and Cynthia Woods Mitchell Institute for Fundamental Physics and Astronomy, and Department of Physics and Astronomy, Texas A\&M University, College Station, TX 77843,  USA\\
$^{41}$Instituci\'o Catalana de Recerca i Estudis Avan\c{c}ats, E-08010 Barcelona, Spain\\
$^{42}$Department of Astronomy, University of California, Berkeley,  501 Campbell Hall, Berkeley, CA 94720, USA\\
$^{43}$Institute of Astronomy, University of Cambridge, Madingley Road, Cambridge CB3 0HA, UK\\
$^{44}$Laborat\'orio Interinstitucional de e-Astronomia - LIneA, Rua Gal. Jos\'e Cristino 77, Rio de Janeiro, RJ - 20921-400, Brazil\\
$^{45}$Department of Astrophysical Sciences, Princeton University, Peyton Hall, Princeton, NJ 08544, USA\\
$^{46}$Department of Physics and Astronomy, University of Pennsylvania, Philadelphia, PA 19104, USA\\
$^{47}$Department of Physics and Astronomy, Pevensey Building, University of Sussex, Brighton, BN1 9QH, UK\\
$^{48}$School of Physics and Astronomy, University of Southampton,  Southampton, SO17 1BJ, UK\\
$^{49}$Computer Science and Mathematics Division, Oak Ridge National Laboratory, Oak Ridge, TN 37831\\
$^{50}$Lawrence Berkeley National Laboratory, 1 Cyclotron Road, Berkeley, CA 94720, USA\\

%%%%%%%%%%%%%%%%%%%%%%%%%%%%%%%%%%%%%%%%%%%%%%%%%%

% Don't change these lines
\bsp	% typesetting comment

\label{lastpage}
\end{document}